\documentclass[aps,onecolumn,superscriptaddress]{revtex4}
\input epsf
\usepackage{epsfig}
\usepackage{amsmath,amssymb,latexsym}
\newcommand{\switchfonts}{\usefont{OT1}{cmr}{m}{it}}
\newcommand{\lie}{\switchfonts \mbox{\symbol{36}} \normalfont}

\newcommand{\nablada}{ \bar{\nabla}_{a} }

\newcommand{\nabladb}{ \bar{\nabla}_{b} }

\newcommand{\nabladc}{ \bar{\nabla}_{c} }

\newcommand{\bbbox}{ \stackrel{ \bar{} }{\Box} }

\newcommand{\tp}{ {t^{\prime}} }
\newcommand{ \kp}{ {k^{\prime}} }
\newcommand{ \kt}{ \tilde{k} }

\newcommand{ \kpp}{ {k^{\prime \prime}} }
\newcommand{ \tpp}{ {t^{\prime \prime}} }

\newcommand{\gb}{\bar{g}}

\newcommand{\sqrtgb}{\sqrt{ -|\gb| } }

\newcommand{\rhobar}{\bar{\rho}}
\newcommand{\pbar}{\bar{p}}
\newcommand{ \ad }{ a^{\dagger} }

\newcommand{ \ak }{ a_{k} }

\newcommand{ \akp }{ a_{k^{\prime}} }

\newcommand{ \adkp }{ a^{\dagger}_{k^{\prime}} }

\newcommand{ \adkpp }{ a^{\dagger}_{k^{\prime \prime}} }

\newcommand{ \akpminusk }{ a_{k^{\prime} - k} }
\newcommand{ \akppminusk }{ a_{k^{\prime \prime} - k} }

\newcommand{ \adkpminusk }{ a^{\dagger}_{k^{\prime} - k}  }
\newcommand{ \adkppminusk }{ a^{\dagger}_{k^{\prime \prime} - k}  }

\newcommand{ \omegak }{\omega_{\vec{k}}}

\newcommand{ \omegakp }{\omega_{\vec{k}^{\prime}}}
\newcommand{ \omegastarkp }{ \omega^{*}_{\vec{k}^{\prime}} }

\newcommand{ \omegakpp }{\omega_{\vec{k}^{\prime  \prime}}}
\newcommand{ \omegastarkpp }{ \omega^{*}_{\vec{k}^{\prime \prime}} }

\newcommand{ \omegakpminusk }{\omega_{(\vec{k}^{\prime} - \vec{k})}}
\newcommand{ \omegastarkpminusk }{ \omega^{*}_{(\vec{k}^{\prime} - \vec{k})} }

\newcommand{ \omegakppminusk }{\omega_{(\vec{k}^{\prime \prime} - \vec{k})}}
\newcommand{ \omegastarkppminusk }{ \omega^{*}_{(\vec{k}^{\prime \prime} - \vec{k})} }

\newcommand{ \psikp}{\psi_{k^{\prime}}}
\newcommand{ \psikpminusk}{ \psi_{(k^{\prime} - k)} }
\newcommand{ \psikpp}{\psi_{k^{\prime \prime}}}
\newcommand{ \psikppminusk}{ \psi_{(k^{\prime \prime} - k)} }

\newcommand{\intkp}{\int_{\Omega_{k^{\prime}}}}
\newcommand{\intkpp}{\int_{\Omega_{k^{\prime \prime}}}}

\begin{document}

\title{ Long-wavelength metric backreactions in slow-roll inflation  }
\author{ B.~Losic }
\affiliation{ Department of Physics \& Astronomy, 
University of British Columbia
6224 Agricultural Road
Vancouver, B.C. V6T 1Z1
Canada }
\author{ W.G.~Unruh }
\affiliation{ Department of Physics \& Astronomy, 
University of British Columbia
6224 Agricultural Road
Vancouver, B.C. V6T 1Z1
Canada }
\affiliation{ Canadian Institute for Advanced Research, Cosmology and Gravitation Program  }
\email{ blosic@physics.ubc.ca   ;  unruh@physics.ubc.ca}

\date{October 13, 2005 }

\begin{abstract}

We examine the importance of second order corrections to linearized cosmological perturbation theory in an inflationary background, taken to be a spatially flat FRW spacetime.
The full second order problem is solved in the sense that we evaluate the effect of the superhorizon second order corrections on the inhomogeneous and homogeneous modes of the 
linearized flucuations. These second order corrections enter in the form of a {\it cumulative} contribution from {\it all} of their Fourier 
modes. In order 
to quantify their physical significance we study their effective equation of state by looking at the perturbed energy density and isotropic pressure to second order. We define 
the energy density (isotropic pressure) 
in terms of the (averaged) eigenvalues associated with timelike (spacelike) eigenvectors of a total stress energy for the metric and matter fluctuations. 
Our work suggests that that for many parameters of slow-roll inflation, the second order contributions to these energy density and pressures may dominate over the first order 
effects for the case of super-Hubble evolution.
These results hold in our choice of first and second order coordinate 
conditions however we also argue that other `reasonable` coordinate conditions do not alter the relative importance of the second order terms.
We find that these second order contributions approximately take the form of a cosmological constant in this coordinate gauge, as found by others 
using effective methods. 

\end{abstract}

\maketitle
\section{Introduction}

The major success of inflationary cosmology is in simultaneously offering an explanation for the homogeneity of the universe along with a mechanism that 
explains its inhomogeneity. The mechanism generating inhomogeneities involves quantum fluctuations in the fields that represent the dominant form of 
stress-energy during the inflationary epoch, which is a postulated 'potential' dominated era of the early universe (which explains the homogeneity). 
The dynamics of the transition from the inflationary era to our current `kinetic` or `rest mass` dominated era are crucial in the formation of inhomogeneities, 
and in particular fix how much these quantum 
fluctuations will be amplified during the transition. In order to fully quantify the effect of these fluctuations on the large-scale geometry of the universe we can in 
principle use the field equations of general relativity to tell us exactly how the coupled matter and metric fluctuations clasically behave. Of course, in practice, we cannot 
solve the full field equations for this scenario so we solve simplified approximations of the field equations within the framework of cosmological perturbation theory. At 
linear
order the quantized cosmological perturbation theory has emerged as the primary tool to investigate the behaviour of fluctuations in the inflationary era,  
where e.g. it predicts an approximately scale-free spectrum of density fluctuations (see \cite{ Mukhanov:1990me} for a comprehensive review). 

However, the beguiling mystery of the cosmological constant and dark matter/energy problems only deepens with increasingly accurate observations. There has been renewed interest
in the past few years in the effect of higher order corrections to the linearized Einstein equations on both early and late-time physics in inflation.  Suggestions have been 
made that superhorizon higher order corrections to linearized theory, on superhorizon scales (i.e. larger than the Hubble radius), can take the form of a negative cosmological 
constant. This could produce a dynamical relaxation 
mechanism for the bare cosmological constant (starting from \cite{Tsamis:1992sx}, to \cite{Geshnizjani:2002wp}, \cite{Kolb:2005me} most recently, and many references therein). 
The measurability or physical reality of these superhorizon backreaction effects has been a contentious issue (see e.g. \cite{Unruh:1998ic}, \cite{Abramo:2001dd} and 
reference therein), and many questions remain regarding the link between local subhorizon physics and these superhorizon backreactions. 

In this paper we do not add to the discussion of measurability, but focus rather on explicitly evaluating the second order corrections to the right hand side of the 
homogeneous Einstein equations
in perturbation theory.   At second order in perturbation theory one might even expect that, since the effect of second order contributions are {\it cumulative} over 
all wavenumbers, their relative amplitude may become comparable to that of first order.Furthermore, if one thinks 
about solving the perturbed Einstein constraint equations for the matter fluctuations and putting these solutions back into the perturbed evolution equations, it is not hard 
to see that some of the second order corrections could in fact be divided by a so-called slow-roll parameter. This only adds to the worry that the second order terms could 
plausibly dominate for a `slow enough` roll in the background. It is clear that in the limit as the slow-roll parameter goes to zero, so that the background universe 
tends to de Sitter, the first order corrections go to zero and the second order fluctuations dominate in their effect on the gravitational field. At what values of the 
slow-roll parameter do the second order perturbations dominate over the linear ones? The most radical possibility is that the slow-roll conditions are precisely the conditions 
that the second order perturbations dominate. Such questions appear to be behind some of the concerns raised by L. Grischuk in \cite{Grishchuk:1994sj} about the consistency of 
linearized perturbation theory in inflation.

The technical complications of sorting out the second order gauge issues and other nonlinear effects such tensor perturbations seeding scalar perturbations are many, but 
rendered tractable with the aid of packages such as GRTensor for Maple \cite{GRTENSOR}. In this paper we calculate the 
cumulative second order contributions to the homogeneous energy density and pressure.  We do not address 
questions of the ultraviolet (short wavelength) regularization of the fluctuations but focus on the superhorizon fluctuations. Indeed, we pay special attention 
to the case where one considers the cumulative effect of Hubble sized to nearly homogeneous contributions on the homogeneous mode. 

In past work effective approximation methods have been used to evaluate such contributions. One popular method characterizes the backreactions in terms of an 
effective energy-momentum tensor $\tau_{a b}$. In this method there are two contributions to $\tau_{ab}$: the quadratic matter energy momentum tensor and the contribution of 
the first order gravity perturbations. Using early work by Brill, Hartle and Isaacson (\cite{BrillHartle:1964}, \cite{Isaacson:1968}) among others, the Einstein 
equations are expanded to second order assuming the linearized equations hold (so that they drop out). Then the remaining terms are spatially averaged with respect to the 
non-dynamical background metric and the resulting equations are interpreted as equations for a new homogeneous metric $\bar{g}_{ab}$ which include the effects of quadratic 
linear perturbations:
\begin{eqnarray}
G_{ab} ( \bar{g}_{ab} ) &=& \kappa ( \bar{T}_{ab} + < \tau_{ab} > ), 
\end{eqnarray}  
where $< \tau_{ab} >$ is the spatially averaged `backreaction` stress-energy defined by 
\begin{eqnarray}
\tau_{ab} &=&  T_{ab} [ (\delta g_{cd})^2, (\delta \phi)^2 ] - \frac{1}{\kappa} G_{ab} [ (\delta g_{cd})^2 ] .
\end{eqnarray}
Here, $\kappa \equiv 8 \pi G$ in units where $c=1$, $\delta^{n}$ indicates the n-th order perturbation of the object it acts on 
(as explained more precisely in the next section). 

It should be noted that, by construction, the zeroth order equations are not obeyed in 
this formulation. In this effective scenario one is solving for a new isotropic background metric which obeys equation (1), and this can 
in itself raise difficult questions of consistency if one is interested in backreactions on inhomogeneous modes (but is ok if one looks at just the homogeneous mode). This 
can be seen by considering the first variation of the Hamiltonian action for a gravitational system, namely
\begin{eqnarray}
\delta H &=& \int ( \delta N \bar{{\cal{H}}}_{\perp} + \delta N_{i} \bar{{\cal{H}}}^{i} + \bar{N} \delta {\cal{H}}_{\perp} + \bar{N}_{i} {\cal{H}}^{i} ) d^{3} x, 
\end{eqnarray}
and assuming that the background constraints do {\it not} hold, i.e. ${\cal{\bar{H}}}_{\perp} \neq 0, {\cal{\bar{H}}}^{i} \neq 0$. If one has {\it homogeneous} variations 
then these linear terms will vanish {\it anyway}, whether or not the background equations of motion are satisfied. This is so because one can 
do a by-parts integration in the first and second terms above whose result will be boundary terms which can then be set to zero under reasonable assumptions \cite{Nieto:2000}. 

Furthermore, it should also be noted that in their effective approach the full second order Einstein equations are not solved, nor are second order coordinate transformations 
considered. This latter fact can be of considerable concern when interpreting the significance of higher order effects \cite{Unruh:1998ic}. In this sense the effective 
approach does not appear, to us,  to be able to convincingly evaluate the higher-order corrections to Einstein's equations, simply because it never actually considers 
them in the context of higher-order perturbation theory. 

Nevertheless, in \cite{Brandenberger:2002sk} and \cite{Mukhanov:1990me} this effective approach is used to evaluate the dominant 
long-wavelength contributions to $\tau_{ab}$. Defining the energy density and pressure at second order by (using $(+,-,-,-)$ as the 
signature) $\delta^{2} \rho\equiv <\tau^{0}_{\ 0}>$ and $\delta^{2} p \equiv - (1/3) <\tau^{i}_{\ i}>$, 
they find that these contributions have the effective equation of state $\delta^{2} p \approx - \delta^{2} \rho$, with 
$\delta^{2} \rho < 0$. This corresponds to the equation of state of a negative cosmological constant. They also find that 
$\delta^{2} \rho$ grows with time, partially because as inflation proceeds more and more length scales exceed the Hubble scale and 
contribute to $\delta^{2} \rho$. These two results combined suggest their main claim, which is that the backreactions 
effectively create a negative, and growing, cosmological constant which can reduce the actual cosmological constant {\it in the large}. 
Locally one might expect the situation to be different. The first order modes should,  {\it locally},  look like simple coordinate 
transformations of the homogeneous solutions, and their effect on higher order metric 
and matter fluctuations to be again that of higher order coordinate transformations.  However, as we stated earlier, we will not focus on these difficult
issues of interpretation in this paper. 

Instead, we focus on evaluating the higher order contributions to the background equation of state, and in particular calculate the 
quantity $\delta^2 \rho + \delta^2 p$ and its dipsersion. We follow a procedure of consistently (though probably not convergently) expanding the Einstein equations to second 
order and solving a subset of them assuming the zeroth and linear order equations hold. We do this about a flat FRW spacetime in which the dominant gravitating 
matter is a slowly-rolling, minimally coupled, scalar field $\bar{\phi}$, and we only study the effect of fluctuations on spatial scales exceeding the 
Hubble radius. In order to define the second order energy density and isotropic pressure, we give an invariant definition of $\delta ^2 \rho$ and $\delta^2 p$ in 
terms of the eigenvalues of the stress energy tensor. Due to the mixing of tensor and scalar waves, these fluctuations will not only arise from second order scalar modes 
but also from quadratic combinations of scalar-scalar and tensor-tensor modes at second order. We find that in general $\delta^2 p + \delta^2 \rho \neq 0$ and  
$\delta^2 \rho < 0$, but that $\frac{ \delta^2 \rho + \delta^2 p }{ \delta^2 \rho}$ does become small.  Perhaps surprisingly, we also find that the relative amplitude of the 
second order dipsersion $< (\delta^2 \rho)^2 >$ dominates over its linear counterpart $< (\delta \rho)^2>$ for a wide range of slow-roll parameters in the background.

\section{ Long-wavelength second order perturbation theory }

In linear cosmological perturbation theory \cite{ Mukhanov:1990me} the tensor, vector, and scalar metric modes all decouple, 
which justifies examining only one class of modes at a time to make the calculations simpler.  Furthermore, depending on the initial 
conditions one takes for the linear metric modes at the beginning of inflation (a contentious issue), the scalar modes are typically 
more important than the gravitational modes at the end of inflation. Given this, the fact that scalar modes directly lead to energy
density fluctuations required for structure formation, and the fact that for longwavelength perturbations gravity wave terms are typically 
suppressed by factors of $k^2/a^2$, 'cosmological perturbations' have become synonymous with scalar perturbations. 
At second order, however, it is well known that pure second order modes have products of tensor, vector, and scalar modes as sources.  
For example, linear tensor fluctuations can induce second order scalar modes via the second order field equations. In order to 
describe this it becomes essential to include all three classes of modes (scalar, vector, and tensor) as sources for the 
higher order gravitational radiation.  

For fluctuations about a spatially flat FRW background in comoving coordinates $(t, \vec{x})$ the 
perturbed metric may be written down as 
\begin{small}
\begin{eqnarray}
%\nonumber
ds^{2} &=& -( 1 + \epsilon A(t, \vec{x}) + \epsilon^2 {\cal{A}}(t, \vec{x}) )dt^2 + 2( \epsilon B_{i}(t, \vec{x}) 
+ \epsilon^2 {\cal{B}}_{i}(t, \vec{x})) dt dx^{i} 
 + a^2(t) ( \delta_{ij} + \epsilon h_{ij}(t, \vec{x}) + \epsilon^2 q_{ij}(t, \vec{x}) ) dx^{i} dx^{j}, 
\end{eqnarray}
\end{small}
where $\epsilon$ is the strength of the linear perturbation,  $i,j,k,...$ denote purely spatial indices, and $a(t)$ is the usual scale factor. 
Together with the matter perturbations of the scalar matter $\bar{\phi}$, defined by  
\begin{eqnarray}
%\nonumber
\phi(t, \vec{x}) &=& \bar{\phi}(t) + \epsilon \Phi(t, \vec{x}) + \epsilon^2 {\cal{F}} (t, \vec{x}),
\end{eqnarray}
the perturbations are $(A, B_{i}, h_{ij}, \Phi)$ at linear order and $({\cal{A}}, {\cal{B}}_{i}, q_{ij}, {\cal{F}})$ at second 
order. The backreactions or higher corrections are suitably integrated quadratic combinations of terms from the former set, and  
they affect the longwavelength part of the latter set of variables.  In 
principle, their evolution is 
determined by substituting the above metric and matter perturbations into the Einstein equations and solving these at second order subject 
to the linearized and background equations, i.e. we solve 
\begin{eqnarray}
\delta^{2} G_{ab} &=& \kappa \left( \delta^{2}  T_{ab}  \right) ,
\end{eqnarray}
and also demand that the linearized and zeroth-order (background) field equations
\begin{eqnarray}
\delta G_{ab} &=& \kappa \left( \delta  T_{ab} \right ) \\
\bar{G}_{ab} &=& \kappa \left( \bar{T}_{ab} \right),
\end{eqnarray}
hold. Here, $G_{ab}$ is the usual Einstein tensor and 
$T_{ab} =  \left\{ \phi_{;a} \phi_{;b} - g_{ab} \left( \frac{1}{2} \phi^{;c} \phi_{;c} + V( \phi ) \right) \right\}$ is the stress-energy 
for a minimally coupled scalar field.  

In the following we assume that the spatial dependence of the fluctuations is of the form $e^{\pm i k_{i} x^{i} }$, and we express the 
longwavelength approximation by 
\begin{eqnarray}
\left( \frac{ k }{a H } \right)^2 \ll 1. 
\end{eqnarray}
Here, $H$ is the Hubble parameter that corresponds to the scale factor expansion $a(t) \sim t^{\alpha}$, $\alpha \gg 1$ and we 
take the potential of the slowly-rolling background scalar field to be 
\begin{eqnarray}
V( \phi ) &=& \Lambda + \beta \phi, 
\end{eqnarray}
for which the only non-trivial slow-roll condition is 
$\frac{1}{\kappa} ( \frac{V_{, \phi}}{V} )^2 =  \frac{ \kappa \beta^{2} }{ H^{4} }  \ll 1$. One should note that $\dot{H} = - \frac{ \kappa \beta^2 }{ 18 H^2}$, and 
we write $\sqrt \alpha \sim \frac{ H }{ \sqrt{ -\dot{H} } } =   \sqrt{ \frac{ 18 H^4 }{ \kappa \beta^2}}$ so that the deSitter limit corresponds to 
$\alpha \rightarrow \infty$. Conservation of energy for a slowly-rolling scalar field in this potential, with initial value $\phi_{0}$, requires that $\phi$ take the form
\begin{eqnarray}
\phi &=& \phi_{0} - \frac{ \beta t }{3 H}, 
\end{eqnarray}
and we shall only consider comoving times $t$ such that $0 \le t \ll \frac{3H\phi_{0}}{\beta}$.

Therefore we have three small parameters in this problem: $\epsilon$ (the strength of the matter and metric fluctuations), 
$\frac{ \kappa \beta^{2} }{ H^4 }$
(slow-roll parameter associated with our choice of inflaton potential), and $\frac{k}{aH}$, the long-wavelength parameter. However, only two of these small parameters 
are independent as the order of the metric and matter fluctuations is a direct product of the physics of the slow-roll parameter, so the scale of $\epsilon$ 
is in some sense dynamically set. Unless otherwise specified, all future references to the order of a quantity will refer to its order in $\epsilon$. 
%The solutions to equations (6) and (7) will form an equivalence class under only a restricted group of diffeomorphisms. One can choose such a restricted 
%diffeomorphism to not only simplify the form of the solutions and the equations themselves, but to also to deduce what part of the metric 
%and matter fluctuations is physical and what part is just a coordinate effect.

The solutions to equations (6) and (7) will be invariant under under a class of diffeomorphisms which are themselves functions of $\epsilon$. By expanding out the
diffeomorphisms order by order in $\epsilon$, one can apply first order, second order, etc., coordinate transformations to the solutions of the first order, second order, 
etc., solutions of (6) and (7). In particular, one can choose these transformations to not only simplify the form of the solutions and the equations themselves, but to 
also deduce what part of the metric and matter fluctuations is physical and what part is just a coordinate effect.

\subsection{ Linear and second order coordinate transformations; gauge fixing } 

As compared to the general covariance of the full (infinite order) theory, which allows for arbitrary coordinate transformations, within 
the framework of second-order perturbation theory we consider only linear and second order infinitesimal parts of these coordinate 
transformations (called gauge transformations in cosmological perturbation theory).  The individual perturbations of the metric or stress-energy 
components will change in some well-defined way under such gauge transformations, and there will in general exist (an infinite number) of combinations 
of fluctuations which are invariant, to second order, under this restricted class of coordinate transformations.

Indeed, one can write a general $\epsilon$-dependent coordinate transformation by
\begin{eqnarray}
\tilde{x}^{\alpha} &=& \tilde{x}^{\alpha}( x, \epsilon ), 
\end{eqnarray}
and define an associated {\it linearized} coordinate transformation by 
\begin{eqnarray}
\zeta^{a} &=& \lim_{\epsilon \rightarrow 0} 
\frac{ \partial \tilde{x}^{c} (x, \epsilon) }{ \partial \epsilon } 
			\frac{ \partial X^{a} (\tilde{x}(x, \epsilon), \epsilon) }{ \partial \tilde{x}^{c} }, 
\end{eqnarray}
where $X^{a} (\tilde{x}(x, \epsilon), \epsilon) = x^{a}, \ \  \forall \epsilon$, and similarly we can define the {\it second-order} 
coordinate transformation by 
\begin{eqnarray}
\chi^{a} &=& \lim_{\epsilon \rightarrow 0} \frac{ \partial }{ \partial \epsilon } 
          \left[ \frac{ \partial \tilde{x}^{c} (x, \epsilon) }{ \partial \epsilon } 
			\frac{ \partial X^{a} (\tilde{x}(x, \epsilon), \epsilon) }{ \partial \tilde{x}^{c} } \right].
\end{eqnarray}
Since the metric $g_{ab}$ is a tensor it will in general transform by 
\begin{eqnarray}
g_{ab} (x, \epsilon) &=& \frac{  \partial \tilde{x}^{c} }{   \partial x^{a}  } \frac{ \partial \tilde{x}^{d} }{   \partial x^{b} } 
										\tilde{g}_{cd}( \tilde{x}(x, \epsilon), \epsilon ), 
\end{eqnarray}
one can show that under a linear coordinate transformation the linear metric fluctuations $\delta g_{ab}$ suffer the 
change (using equation (13)) 
\begin{eqnarray}
 \lim_{\epsilon \rightarrow 0} \frac{ \partial g_{ab} (x, \epsilon ) }{ \partial \epsilon } & \equiv & \delta g_{ab} 
= \delta \tilde{g}_{ab} + \lie_{\zeta} ( g_{ab}(\epsilon = 0) ) \equiv \tilde{g}_{ab} + \lie_{\zeta} \bar{g}_{ab},
\end{eqnarray}
where $\lie_{\zeta}$ is the usual Lie derivative along the vector $\zeta^{a}$. Similarly using equations (13) and (14) and the notation of 
equation (16),  it is not hard to show that 
\begin{eqnarray}
\delta^{2} g_{ab} &=& \delta^{2} \tilde{g}^{\prime}_{ab} + ( \lie_{\zeta}^{2} + \lie_{\chi} ) \bar{g}_{ab} 
						+ 2 \lie_{\zeta} \delta \tilde{g}_{ab}, 
\end{eqnarray}
so that in particular one can see that the change in (gauge transformation of) the second order fluctuation $\delta^{2} g_{ab}$ depends on $\zeta^{a}$, i.e. it depends on the 
linearized coordinate transformation $\zeta^{a}$ as well as the  second order transformation $\chi^{a}$. This 
implies, for example, that the changes in the second order stress energy $\delta^{2} T_{ab}$ of equation (6) caused by the gauge transformation will  
depend on both $\zeta^{a}$ and $\chi^{a}$, the first and second order gauge transformations. 

In standard cosmological perturbation theory one usually makes a particular 
choice of $\zeta^{a}$ to simplify the interpretation of the fluctuations, e.g. the longitudinal gauge \cite{ Mukhanov:1990me}: 
\begin{eqnarray}
\zeta^{0} &=& B - a \dot{E}, \\
\zeta^{i} &=& -\partial^{i} E, 
\end{eqnarray}
so that $\delta g_{0i} = 0$ and $\delta g_{ij} \propto \bar{g}_{ij}$. 
It should be noted that longitudinal gauge only fixes the {\it scalar} part of the metric into diagonal form, and its simple form relies 
crucially on the form of the anisotropies of the perturbed stress energy.  However for our problem, since we obviously cannot diagonalize the {\it entire} linear 
order metric because of the presence of TT gravity waves, we find it convenient to emulate the harmonic gauge from the full theory, namely
\begin{eqnarray}
\frac{1}{\sqrt{-|g|}} \partial_{a} \left( \sqrt{-|g|} g^{ab} \right) = 0
\end{eqnarray}
To linear order this gauge choice corresponds to setting
\begin{eqnarray}
B^{i}_{,i} &=& 0, \\
\partial^{j} \left( h_{ij} - \frac{\delta^{\ell m} h_{\ell m} }{3} \delta_{ij} \right) &=& 0,
\end{eqnarray}
which fixes, to within trivial residual gauge freedoms, the linearized metric perturbation regardless of the form 
of the perturbed stress energy and is sometimes known as the Poisson gauge \cite{Bertschinger:1993xt}.  Clearly the longitudinal gauge for the scalar sector is a 
special case of (21) and (22), since it restricts only the {\it potentials} $B, E$ of the metric fluctuations $B_{,i} , E_{|ij}$.  The primary 
physical advantage in using this generalization of the longitudinal gauge is that one can unambiguously transform to any gauge while easily keeping track of 
the residual freedoms, while the primary mathematical advantage is that they lead 
to a compact form for the perturbed hamiltonian and momentum constraints. One can always transform from this gauge to any other gauge since the 
transformations are {\it algebraic} in nature (as opposed to, say, the nonlocal integrals that take one to synchronous gauge) (see 
\cite{Bertschinger:1993xt} for more details).

\section{ Total energy density and pressure at second order }

How do the classical metric and matter fluctuations at second order influence the background equation of state, and in particular how does one assess the
influence of the gravitational backreactions? 
Fortunately, within finite order perturbation theory we can avoid the conceptual and technical problems involved in defining 
local or quasi-local definitions of gravitational energy-momentum because, by definition, we have a preferred decomposition of the 
spacetime metric. This allows us to exclusively attribute 'energy' to the 'dynamical part' ($ \delta g_{ab}, \delta^{2} g_{ab} )$ of the metric as opposed to the 
'background part' ($\bar{g}_{ab}$) simply because there exists 
a 'background derivative' ($\nablada$) against which to measure any such `dynamics`. Thus we can define a 'relative gravitational stress energy' $\tau_{ab}$ 
of the fluctuations with respect to the curved background, and in particular we can take combinations of $\tau_{ab}$ and the stress energy $T_{ab}$ to study the fluctuations 
in the pressure and energy density at second order.

A general formalism to define conserved quantities and conservation laws with respect to curved background spacetimes has already been developed by 
Katz, Bicak, and Lynden Bell in \cite{Katz:1996nr}.  The basic idea is to start from the Lagrangian
\begin{eqnarray}
L(\epsilon) &=& \frac{\sqrt{ -|g|(\epsilon) }}{2 \kappa } \left[ C^{\rho}_{\mu \nu} (\epsilon) C^{\sigma}_{\rho \sigma} (\epsilon) - C^{\rho}_{\mu \sigma}  (\epsilon)
														C^{\sigma}_{\rho \nu}  (\epsilon) \right]
			- \frac{1}{2 \kappa } \left( g^{\mu \nu}  (\epsilon) \sqrt{ -|g|(\epsilon) }  - \gb^{\mu \nu} \sqrtgb \right) \bar{R}_{\mu \nu} + L_{M}, 
\end{eqnarray}
where $2 C_{abc} (\epsilon) \equiv 2 (  \Gamma_{abc} (\epsilon) -  \bar{\Gamma}_{abc} ) 
			=  ( \nabladb g_{ca} (\epsilon)  + \nabladc g_{ba} (\epsilon) - \nablada g_{bc} (\epsilon) )$ and $L_{M}$ is the `matter` Lagrangian. Note 
that $ \bar{C}^{a}_{bc} = 0$, 
so that in the background $\bar{L} = 0$. From $L$ one can build vector densities $I^{\mu}$ which are conserved in the sense that $\bar{\nabla}_{\mu} I^{\mu} = 0$. We refer the 
reader to \cite{Katz:1996nr} for the details of their construction and simply quote the result 
\begin{eqnarray}
I^{\mu} &=& \left[  \sqrt{ -|g|(\epsilon) } T^{\mu}_{\nu} (\epsilon) - \sqrtgb \bar{T}^{\mu}_{\nu} 
+ \frac{1}{2}  \left( g^{\rho \sigma}  (\epsilon) \sqrt{ -|g|(\epsilon) }  - \gb^{\rho \sigma} \sqrtgb \right)  \bar{R}_{\rho \sigma} \delta^{\mu}_{\nu} 
		 + \sqrt{ -|g|(\epsilon) } t^{\mu}_{\nu} \right] \zeta^{\nu} \\
\nonumber
&& + \sqrt{ -|g|(\epsilon) } \left( \sigma^{\mu [\rho \sigma]} \partial_{[\rho} \zeta_{\sigma]} + {\cal{Z}}^{\mu} (\zeta^{\nu})  \right) \\
\nonumber
&\equiv&  \left[  \sqrt{ -|g|(\epsilon) } T^{\mu}_{\nu} (\epsilon) - \sqrtgb \bar{T}^{\mu}_{\nu} 
+ \sqrt{ -|g|(\epsilon) } \tau^{\mu}_{\nu}   \right] \zeta^{\nu} 
		+ \sqrt{ -|g|(\epsilon) } \left( \sigma^{\mu [\rho \sigma]} \partial_{[\rho} \zeta_{\sigma]} + {\cal{Z}}^{\mu} (\zeta^{\nu})  \right)
\end{eqnarray}
where $\zeta^{\mu}$ is the arbitrary, smooth, vector field and $\tau^{a}_{b}$ is the analogue of the 
Einstein pseudotensor (defined in terms of $C^{a}_{bc}$ instead of $\Gamma^{a}_{bc}$).  The first term on the right hand side of (24), in square brackets,
can be interpreted as the relative stress energy of the fluctuations with respect to a given background if we expand it to the desired order in $\epsilon$. The last group of 
terms can be interpreted in terms of the relative helicity of the perturbations with respect to the background  \cite{Katz:1996nr} and we shall not consider these in 
this work. We denote the gravitational parts of the relative stress energy by $\tau_{ab}$, which is only a tensor to second order in $\epsilon$. 

\subsection{ Eigenvalues of the total stress energy } 

One often defines (the rotationally invariant but not boost invariant) energy density and isotropic pressure of a perfect fluid by
$-\rho \equiv g^{00} T_{00}$ and $3p \equiv g^{ii} T_{ii}$.  While this is relatively straightforward to interpret and implement in linearized perturbation theory,
at second order one gets complications such as having to subtract off shear (offdiagonal) stresses $\sim \delta p_{i} \delta p_{j}, i \neq j,$ from the diagonal isotropic 
contributions $\sim ( \delta p_{i} )^2$. From this and other points of view it turns out to be extremely useful to consider the eigenvalues of the mixed-valence total 
stress energy of the fluctuations, i.e. to consider the eigenvalues of the tensor
\begin{eqnarray}
\bar{T}^{a}_{\ b} + \delta T^{a}_{\ b} + \delta^2 T^{a}_{\ b} + \delta^2 \tau^{a}_{\ b}, 
\end{eqnarray}
where the last term $\delta^2 \tau^{a}_{\ b}$ is the second order part of the relative gravitational stress energy, $\tau_{ab}$, described above.
Since it is of mixed valence, it transforms from coordinates $x$ to $\bar{x}$ as $( \partial x / \partial \bar{x}) ( \partial \bar{x} / \partial x )$ and 
therefore has gauge-covariant eigenvalues $\lambda_{i}$ associated to timelike and spacelike eigenvectors, which are calculated by solving the equation
\begin{eqnarray}
det \left( \bar{T}^{a}_{\ b} + \delta T^{a}_{\ b} + \delta^2 T^{a}_{\ b} + \delta^2 \tau^{a}_{\ b} - \lambda_{i} \delta^{a}_{\ b} \right) = 0
\end{eqnarray}
At zeroth order, since we are not in deSitter but in a slow-roll spacetime, the one timelike eigenvalue and three spacelike eigenvalues (associated with their 
respective eigenvectors) are different and are in that sense 'sufficiently separated'; this 
is sufficient to guarantee that the perturbations obey this property of being well-separated as well (the three spatial eigenvalues are not, but we only are interested in 
their average value). We define the energy density as minus the eigenvalue of the timelike 
eigenvector and the cumulative isotropic pressure as the average of the distinct eigenvalues associated with their respective spacelike eigenvectors. We also emphasize that 
the eigenvectors at second order will in general point in different directions, but that all that matters in our calculations is the averaged contribution obtained 
after quantum averaging over $k$. After such averaging the terms like $\delta \rho(k) \delta \rho(k^{\prime})$ collapse to the diagonal terms $( \delta \rho(k) )^2$. 

We can express these eigenvalues in terms of scalars formed from the stress tensor and powers thereof. For example, to linear order one may find the averaged 
eigenvalues $\delta \rho$ and 
$\sum_{i} \delta p_{i}$ by perturbing the expressions
\begin{eqnarray}
T^{a}_{\ a} &=& -\rho + \sum_{i} p_{i} \\
S^{a}_{\ b} S^{b}_{\ a} &=& \frac{3}{4} \left[ \rho^2 + \frac{2 \rho}{3} \sum_{i} p_{i} + \frac{1}{3} \left( 4 \sum_{i} ( p_{i} )^2 
												- \left( \sum_{\ell} p_{\ell} \right)^2 \right) \right], 
\end{eqnarray}
where $S_{ab} \equiv T_{ab} - \frac{T^{m}_{\ m}}{4} g_{ab}$. To linear order, these relations are equivalent to 
\begin{eqnarray}
\delta (T^{a}_{\ a}) &=& -\delta \rho + \sum_{i} \delta p_{i} \\
\frac{4}{3} \delta (S^{a}_{\ b} S^{b}_{\ a} ) &=& 2 (\rhobar + \pbar)  (\delta \rho + \frac{1}{3} \sum_{i} \delta p_{i}),
\end{eqnarray}
where $\rhobar + \pbar = \frac{ \beta^2 }{9 H^2}$. 
Substituting the explicit expressions in terms of the metric and matter fluctuations for the left hand side, one 
finally obtains the desired expressions for the energy density and cumulative isotropic pressure. To higher order this procedure becomes more complicated since, as we 
mentioned above, pressure contributions like $(\sum_{i} p_{i} )^2$ contain both diagonal contributions like $\sum_{i} p_{i}^2$ and offdiagonal shear contributions like 
$\sum_{i \neq j} \delta p_{i} \delta p_{j}$. However the basic strategy is similar and in any case equivalent to solving the above determinant 
using scalars such as the trace, `double trace`, and determinant, as we now sketch out. 

For the second order case, the starting point is 
\begin{eqnarray}
\delta^2 (T^{a}_{\ a} + \tau^{a}_{\ a}) &=& -\delta^2 \rho + \sum_{\i} \delta^2 p_{i} \\
\delta^2 (\frac{4}{3}S^{a}_{\ b} S^{b}_{\ a} ) &=& ( \delta \rho )^2 + \left( \sum_{i} \delta p_{i} \right)^2 + \frac{2}{3} \delta \rho \sum_{i} \delta p_{i} 
											- \frac{8}{3} \sum_{i \neq j} \delta p_{i} \delta p_{j} 
									+ 2 (\rhobar + \pbar)  (\delta^2 \rho + \frac{1}{3} \sum_{i} \delta^2 p_{i}), 
\end{eqnarray}
where $S_{ab} \equiv ( T_{ab} + \tau_{ab} ) - \frac{T^{m}_{\ m} + \tau^{m}_{\ m}}{4} g_{ab}$. The appearance of terms like  $\sum_{i \neq j} \delta p_{i} \delta p_{j}$
and related cross-terms complicates the isolation of the desired eigenvalues $\sum_{i} \delta^{2} p_{i}$. In order to eliminate such terms we consider the second order 
perturbation of the cube of the trace-free part of the total stress energy. Combined with equations (31) and (32), this will give 
us another equation and with it the possibilty of cancelling these shear terms in terms of some function of metric and matter fluctuations. The general expression for 
the cube is 
\begin{eqnarray}
-\frac{8}{3} S^{a}_{\ b} S^{c}_{\ a} S^{b}_{\ c}  &=& \rho^3 - \sum_{i} p_{i}^3 - \rho \sum_{i} p_{i}^2 + \sum_{\ell \neq m} p_{\ell} p_{m}^2 + \rho^2 \sum_{i} p_{i}
								+ 2 \rho \sum_{i \neq j} p_{i} p_{j} - 2p_{1} p_{2} p_{3}
\end{eqnarray}
which to second order is 
\begin{small}
\begin{eqnarray}
\nonumber
\delta^2 \left( -\frac{8}{3} S^{a}_{\ b} S^{c}_{\ a} S^{b}_{\ c} \right) &=&  
		 ( \rhobar + \pbar)  \left( 3( \delta \rho )^2 - \left( \sum_{i} \delta p_{i} \right)^2 + 2 \delta \rho \sum_{i} \delta p_{i} 
											+ 4 \sum_{i \neq j} \delta p_{i} \delta p_{j} \right)
			+ 3 \left( \rhobar + \pbar  \right)^2 \left(  \delta^2 \rho + \frac{1}{3} \sum_{i} \delta^2 p_{i} \right), 
\end{eqnarray}
\end{small}
and which in turn has the right form to solve for the shear terms we are desiring to eliminate from expression (30). Substituting in the expression of 
$\delta^2 (-\frac{8}{3} S^{a}_{\ b} S^{c}_{\ a} S^{b}_{\ c} )$ in terms of metric and matter fluctuations and solving for the four averaged 
eigenvalues $\delta \rho, \delta^2 \rho, \sum_{i} \delta p_{i}, \sum_{i} \delta^2 p_{i}$ in terms of the mertic fluctuations, we finally get 
\begin{equation}
\left . 
\begin{array}{rcl}
4  (\rhobar + \pbar)  ( \delta^2 \rho + \frac{1}{3} \sum_{i} \delta^2 p_{i} ) 
+ \left[ \frac{1}{3} \left( \sum_{i} \delta p_{i} \right)^2 + 3 (\delta \rho^2) + 2 \delta \rho \sum_{i} \delta p_{i} \right]
&=& \delta^{2} {\cal{\vartheta}} + \frac{2}{3 (\rhobar + \pbar)} \delta^{2} \Theta  \\
-\delta^2 \rho + \sum_{i} \delta^2 p_{i} &=& \delta^{2} {\cal{T}} \\
2 (\rhobar + \pbar) ( \delta \rho + \frac{1}{3} \sum_{i} \delta p_{i} ) &=& \delta {\cal{\vartheta}} \\
- \delta \rho + \sum_{i} \delta p_{i} &=& \delta {\cal{T}}
\end{array} \right\},
\end{equation}
where $\theta \equiv \frac{4}{3}S^{a}_{\ b} S^{b}_{\ a}, \Theta \equiv -\frac{8}{3} S^{a}_{\ b} S^{c}_{\ a} S^{b}_{\ c}, {\cal{T}} = g^{ab} ( T_{ab} + \tau_{ab} )$. 
The simultaneous solutions to these two sets of coupled equations are the (averaged) eigenvalues one would find directly from the matrix represented by the total 
stress-energy, expression (25). They are 
\begin{equation}
\left . 
\begin{array}{rcl}
\delta p &\equiv& \frac{1}{3} \sum_{i} \delta p_{i} =  \frac{ \delta {\cal{T}} }{4} +\frac{1}{8(\rhobar+ \pbar)}  \delta \theta \\
\delta p + \delta \rho &=& \frac{1}{2 (\rhobar + \pbar)} \delta \theta \\
\delta^2 p &\equiv& \frac{1}{3} \sum_{i} \delta^2 p_{i} = \frac{1}{\rhobar + \pbar} 
\left[ \frac{ \delta^2 \theta }{16} +  \frac{\delta^2 \Theta}{24(\rhobar+\pbar)}  - \frac{ 3 ( \delta \theta )^2 }{64 (\rhobar+\pbar)^2} \right] 
															+ \frac{ \delta^2 {\cal{T}} }{4} \\
\delta^2 \rho + \delta^2 p &=& \frac{1}{\rhobar+\pbar} 
\left[  \frac{ \delta^2 \theta }{4} +  \frac{ \delta^2 \Theta }{6(\rhobar+\pbar) } - \frac{ 3 ( \delta \theta )^2 }{16 (\rhobar+\pbar)^2} \right]
\end{array} \right \}
\end{equation}

\subsubsection{ Linear contributions to the energy density and pressure }

At linear order the two values $\delta \rho, \sum_{i} \delta p_{i}$ of the total stress energy $\delta T^{a}_{\ b}$, comprised of only the stress energy of matter, 
can easily be found. Assuming the longitudinal gauge-fixing, we find
\begin{equation}
\left . 
\begin{array}{rcl}
\delta p &=& - \frac{ \beta }{3 H} \left( \partial_{t} - 3 H \right) \Phi - \frac{ \beta^2}{18 H^2} A \\
\delta \rho + \delta p  &=&  2 \beta \Phi  \\
\end{array} \right\},
\end{equation}

To linear order, only scalar modes can induce energy density and pressure fluctuations. In the longitudinal gauge fixing specified by equations (18),(19), the (constrained) 
equation of motion for the spatial diagonal metric perturbation $\psi$ in the long-wavelength limit, assuming slow-roll, is simply
\begin{eqnarray}
(\partial^2_{t} + H \partial_{t} )\psi(t) &=& 0, 
\end{eqnarray}
whose nondecaying solution can be taken to be a nonzero constant ( $\equiv \psi$ ). This is to be distinguished from the pure deSitter case where this constant is precisely 
zero, since there are no physical linear scalar modes in pure dS \footnote{ One can set up a scalar field with intial velocity on a flat potential, which will rapidly 
decelerate due to Hubble friction and drive the spacetime to dS. During this deceleration, which will violate slow-roll,  there will of course be scalar gravitational 
perturbations but the physical (noncoordinate) part of this radiation will smoothly tend to zero as $\dot{\phi} \rightarrow 0$. In pure de Sitter the statement that the only 
nongauge excitations of the metric are TT is true on all physical length scales.}. The corresponding matter perturbation 
$\Phi$ is easily found via the constraint equations (namely $\Phi = ( 3H^2/\beta ) \psi$ ). Using this and the constraint equations from (7) to express the result in 
terms of $\psi$, we find the dominant contributions to $\delta p$ and $\delta \rho$ are 
\begin{eqnarray}
\delta p &\approx& - \frac{3 \alpha^2 }{ \kappa t^2 } \psi \stackrel{\alpha \rightarrow \infty}{=}  - \frac{3 H^2 }{ \kappa } \psi \\
\delta \rho + \delta p &\approx& \frac{\psi H^2}{9 \kappa} \left(  54 \epsilon_{LW} - (6Ht -1) \epsilon_{SR}  \right), 
\end{eqnarray}
where $ \left( \frac{k}{aH} \right)^2 \equiv \epsilon_{LW}, \frac{ \kappa \beta^2 }{H^4} \equiv \epsilon_{SR}$. Although the right hand of the latter equation is in some 
sense small, it is not zero. Therefore the linearized contribution to the equation of state, though highly suppressed, will still 
depend on the details of the small parameters. This point may perhaps be more obvious if one considers how the time evolution equation for the scalar field fluctuations 
( during slow-roll ) is modified by the inclusion of the metric fluctuations, namely
\begin{eqnarray}
\ddot{\Phi} + 3H \left( 1 - \frac{1}{ 9 (\frac{\kappa( \phi_{0} - \phi)}{\partial_{\phi} ln (V(\phi))} - 9\frac{\epsilon_{LW}}{\epsilon_{SR}} ) } \right) \dot{\Phi}
	+  3 H^2 \epsilon_{LW} \left( 1 - \frac{2}{27} \frac{ \epsilon_{SR} }{ \epsilon_{LW}} \right) \Phi &=& 0,\ \  0 \le t \ll \frac{3H \phi_{0}}{\beta} 
\end{eqnarray}
This shows how the gravitational fluctuations, at the linearized level,  effectively induce a negative effective mass for the fluctuating scalar field 
as well as modify the effective Hubble parameter - all in a way which depends on what values we take for $\epsilon_{SR}, \epsilon_{LW}$.  Since the longitudinal gauge 
(or equivalently, a longitudinal choice of gauge invariant variables) admits no residual linearized coordinate tranformations, these fluctuations cannot be associated with 
coordinate modes and are hence physical.  

It is also important to note that the right hand side of (36) contributes at a given, fixed, wavenumber $k$ to the linear fluctuations $\delta \rho$, 
$\delta p$. At higher order we generically expect that contributions to the nearly homogeneous {\it second} order modes of $\delta^2 \rho$ and $\delta^2 p$ 
will be cumulative over a broad range of $k$ of the {\it linear} modes. It is this enhanced, cumulative, contribution to the second order modes that can make 
the nonlinear contributions to the equation of state nontrivial.  

\subsubsection{ Second order energy density and pressure perturbations at fixed $k$ }

Considering for a moment the situation at second order with fixed $k$, the partially gauge fixed eigenvalues of the total stress-energy are, for the generic case with no 
gauge-fixing,
\begin{eqnarray}
\nonumber
\delta^{2} p &=&  - \frac{ \beta }{3 H} \left( \partial_{t} + 3 H \right) {\cal{F}} + \frac{ \beta^2}{18 H^2} {\cal{A}} 
	      + \frac{1}{2} ( \partial_{t} \Phi )^2  - \frac{12H^2}{32 \kappa } \psi^2 - \frac{3 H^2}{32 \kappa} C^2 \\
&& +  \frac{3k^2}{2a^2} \Phi^2	
			+ \frac{ \beta \Phi}{3Ha^2} k_{i} B^{i} 
+ \frac{\beta^2}{18 H^2 } \left( -\frac{B_{i} B^{i}}{a^2} +  A^2 \right) 
						+ \frac{\beta}{3H}  A \partial_{t} \Phi   \\
%\nonumber
\delta^{2} \rho &=&  -\delta^{2} p + 2 \beta {\cal{F}} - \frac{93H^2}{4 \kappa } \psi^2 + \frac{93 H^2}{\kappa} C^2,
\end{eqnarray}
where $C^2$ is the squared amplitude of the linear tensor fluctautions. As in the linear case, the matter fluctuations ${\cal{F}}$ are related to the metric fluctuations 
via the second order constraints from (6). The behaviour of the second order scalar perturbations will be influenced by not only the constant scalar modes at linear order, 
but also the linear tensor-tensor terms.  One would thus expect that the scalar modes at second order will  
become time dependent. In fact, if we pick the `harmonic-analogue` gauge of Section I, so that at linear and second 
order the scalar sectors of the metric are diagonal, we see
\begin{small}
\begin{eqnarray}
t &\rightarrow& t + \left(  B - a \dot{E} \right) \\
\nonumber
&& + \left[   {\cal{B}} - a {}^{(2)} \dot{E} + 3E(3B - 10a\dot{E})k^2 
+ 4( a^2 \ddot{E} + H \dot{E} ) ( 2a \dot{E} - B ) + 2 a \dot{E} (A - \frac{5}{2} a\dot{B} )  
										+ B ( a \dot{B} - 2A ) -2aE \dot{\psi} \right] \\
x^{i} &\rightarrow& x^{i} 
+ \partial^{i} \left\{  -E  
	+ \left[  - {}^{(2)} E - \frac{33}{2} k^2 E^2 - 2\psi E + \frac{9}{2}  a^2{ \dot{E} }^2 - \frac{1}{2} B^2 
										- 4a \dot{E} B \right] \right\} 
\end{eqnarray}
\end{small}
then we obtain, using the spatial second order scalar field equations in this new coordinate system,
\begin{eqnarray}
{\cal{A}} &=& - {\cal{Q}} +  \underbrace{ \frac{2}{3} \left[ (  h_{+} )^2 + h_{+} h_{-} +(  h_{-} )^2 \right] }_{\mbox{TT-TT sector}}
					+ \underbrace{ \psi^2 - 2 \kappa \Phi^2 }_{\mbox{ Scalar-Scalar sector}}  , \\
{\cal{B}} &=& 0,  \\
{}^{(2)} E&=& 0,
\end{eqnarray}
where $q_{ij} \equiv {\cal{Q}} \delta_{ij} +  (\partial_{i} \partial_{i} - \delta_{ij} \frac{\bar{\Delta}}{3} ) {}^{(2)} E$ and 
$h_{+}, h_{-} \in \Re$ denote the two TT independent degrees of freedom of $h_{ij}$.  Note that the second-order lapse ${\cal{A}}$ contains contributions from the 
TT-TT gravitational wave contributions at linear order. In effect this is the longitudinal gauge at second 
order. It can easily 
be shown that it admits no residual scalar coordinate freedoms to second order. Within this gauge, under the slow-roll and longwavelength approximations and using the 
constraints to express everything in terms of the metric fluctuations, the equations of motion for ${\cal{Q}}$ are 
\begin{eqnarray}
(\partial^2_{t} + H \partial_{t} ) {\cal{Q}}(t) &=& 
			H^2  \left[ 24Ht -162\frac{ k^2 }{a^2 H^2 } \left( \frac{H^4}{\kappa \beta^2} \right) \right] \psi^2
			+ \left[ \frac{70 k^2 }{3a^2} -  \frac{2 \kappa \beta^2 t}{9 H} \right] C^2,    
\end{eqnarray}
where $C^2 \equiv \left(  ( h_{+} )^2 + h_{+} h_{-} +(  h_{-} )^2 \right) \in \Re$. The growing solution for ${\cal{Q}}(t)$ is, for a fixed $k$ mode of $\psi$, $C$, 
and assuming that $a \sim a_{0} t^{\alpha}$ with $\alpha >> 1$, is 
\begin{eqnarray}
{\cal{Q}}(t) &\approx& \left\{ 24 \alpha^2 \ln(t)   
- \frac{81k^2 \alpha^2}{ \kappa \beta^2} \left( \frac{ t^{-2\alpha} }{ a_{0}^2} \right) \right\} \psi^2 
+  \frac{35k^2}{3 \alpha^2  } \left( \frac{  t^{-2\alpha } }{ a_{0}^2}  \right) C^2 + D, 
\end{eqnarray}
where $D$ is a  constant of integration deduced by the initial conditions of ${\cal{Q}}$, implicitly set to zero by analogy with the linearized sector. This shows that the 
effect of a given linearized mode at some $k$ on the nearly homgeneous mode of ${\cal{Q}}(t)$ is to make it time-dependent, as expected.

When we compute $\delta^{2} p$ and $\delta^{2} \rho$ in our exhaustive coordinate system we use the second and linear order constraints to eliminate the linearized matter 
fluctuation $\Phi$ and the second order matter fluctuation ${\cal{F}}$, so that the final expression is left only in terms of the metric perturbations. 
Following such a procedure, we obtain the following result for the dominant longwavelength contributions: 
\begin{eqnarray}
\delta^{2} p &\approx& -\frac{3H(t)}{\kappa} \left( H(t) + \partial_{t}  \right) {\cal{Q}}(t) - \frac{54 H(t)^6 \psi^2}{ \kappa^2 \beta^2} 
													                                - \frac{3 H^2 C^2}{16\kappa }   \\
\delta^{2} p + \delta^{2} \rho &\approx&  
			+ \frac{216 \psi^2 H(t)^4 k^2 }{a^2 \kappa^2 \beta^2 } + \frac{45 H^2 C^2}{4\kappa }
\end{eqnarray}
At first glance, it is immediately apparent that if we set $C^2 = 0, {\cal{Q}}(t) = 0$ (as Brandenberger et al do in \cite{Brandenberger:2002sk} and references therein) 
then we obtain $\delta^{2} \rho < 0$ but that $\delta^{2} p + \delta^{2} \rho $ is not in general zero because of terms that involve ratios of small parameters:
\begin{eqnarray}
\delta^{2} \rho &\approx&  - 54 \frac{\alpha^6 \psi^2}{\kappa^2 t^6 \beta^2}  \\
\delta^{2} p + \delta^{2} \rho &\approx& 
			+ 216 \frac{k^2 \alpha^4 \psi^2}{a_{0}^2 t^{2\alpha}  t^4 \kappa^2 \beta^2},
\end{eqnarray}
However, we do find that, given $C^2 = 0, {\cal{Q}}(t) = 0$,  quadratic scalar-scalar backreaction contributions generally do 
mimic that of a negative cosmological constant since $( \delta^2 p + \delta^2 \rho)/\delta^2\rho \sim (k/aH)^2$, which is set to zero in their analysis.
Nevertheless, for sufficiently slow-roll it seems worrisome that for Hubble-scale size fluctuations the second order contributions $\delta^{2} p$ and 
$\delta^{2} \rho$ can be comparable to $\delta \rho$ and $\delta p$ at a given $k$, since as they stand equations (52) and (53) show that the second order 
contributions carry 'extra factors' of slow-roll enhancement compared to equations (38) and (39). These extra factors come directly from solving the constraints for the 
matter variables\footnote{ Solving them for the matter variables also introduces the same factors, of course, but it is more subtle to take the longwavelength limit in 
slow-roll.}.

However, it is clearly inconsistent to set ${\cal{Q}}(t)$ to zero in our approach since it is also of second order and in fact contains 
terms proportional to $\psi^2$. Such a tactic also violates the second order field equations, which take into account that the  
metric fluctuation ${\cal{Q}}(t)$ at a given scale will receive contributions from all of the fourier linear modes under consideration 
(in our case, as we explain below, from the horizon to an approxiamtely homogeneous cutoff).  Furthermore since the purely second order contributions to the energy 
density and pressure at these superhorizon scales are not constants but can evolve in comoving time, the contributions may change significantly during the slow-roll era to the 
end of inflation.

In the following section we will examine the cumulative effect of the quadratic combinations of linear modes, averaged over superhubble scales, onto a given scale of the 
purely second order modes. For simplicity we will initially focus on the effect on the second order homogeneous mode. 
We strongly emphasize that the main gauge-fixing of the paper, as 
described by relations (21) and (22), is not well-defined in the strictly homogeneous limit $k = 0$.  An appropriate gauge transformation must be made to sensibly take the 
homogeneous limit and make these manipulations, and we discuss this in the next section. 

\subsubsection{ IR (super-Hubble) contributions from the backreactions }

We now consider the cumulative contributions to the energy density and pressure at second order due to the superhorizon modes. 
Considering only super Hubble fluctuations we know that the dominant linear modes 
are independent of time since the background equation of state, during slow-roll, is approximately time-independent. In terms of the Fourier-decomposed 
$\psi_{k}$, the quantum fluctuations (which we take to be Gaussian) during this era depend on $k$ in such a way that the fluctuations per decade are a constant. Although 
strictly speaking 
cosmological fluctuations are quantized in terms of a reduced variable such as e.g. the Mukhanov-Sasaki (MS) variable $\nu = a( \Phi - \frac{\beta}{3H^2} \psi)$, one can 
always use the linearized constraints to simply relate (in the longwavelength limit) $\nu$ and $\psi$ up to time-dependent factors (see \cite{ Mukhanov:1990me} 
for more details). Indeed, the spatial two-point correlation function of $\psi$ is (after an angular integration)
\begin{eqnarray}
<0| \hat{\psi}(t,\vec{x}) \hat{\psi}(t,\vec{x} + \vec{r}) | 0> \equiv < \psi^2 >  
		= \int_{k_{min}}^{aH} \frac{dk}{k} \frac{sin(kr)}{kr} \left[ \frac{k^3}{4 \pi^2} |\psi_{k}(t)|^2 \dot{\bar{\phi}}^2   \right],
\end{eqnarray}
where $\hat{\psi}(t,\vec{x})$ is the quantum operator associated with $\psi$, expanded in the classical basis of plane waves. The Fourier transform of the two-point function 
is the power spectrum, and completely characterizes Gaussian fluctuations in the sense that all higher correlation functions can be expressed in terms of it. 
Here, $|0>$ is the vacuum chosen so that 
the modes of the reduced MS variable $\nu_{k}$ obey $\nu_{k}(t_{0}) \sim k^{-1/2}, \dot{\nu_{k}} \sim i k^{1/2}$ at some initial time $t_{0}$, which in turns implies a set 
of more complicated conditions on $\psi_{k}$ which are not illuminating at this stage ( see \cite{ Mukhanov:1990me} for more details ). The metric fluctuations $<\psi^2>$
at the horizon scale are related to the density contrast fluctuations $<(\delta \rho / \rhobar)^2>$ by equations (38) and (39), and one can easily show that 
$<(\delta \rho)^2> = 36 (H^4 / \kappa^2) <\psi^2>$. Using all of this, it is relatively straightforward to show that 
\begin{eqnarray}
     k^3 |\psi_{k}|^2 = \frac{1}{4} \frac{H^4}{(2 \pi \dot{\bar{\phi}})^2} = \frac{1}{4} \frac{9 \kappa}{\epsilon_{SR}} \left( \frac{H}{2 \pi} \right)^2, 
\end{eqnarray} 
( see  \cite{Martin:2003kp} for more details ). We take this to hold to some almost homogeneous scale, say $k = k_{min} << aH$, or in other words we cut off the 
infrared divergence of the linearized fluctuations at some scale $k = k_{min}$, so that for $k \sim 0 , |\psi_{k}|^2 \sim 0$. The factors of $4$ on the right hand 
side come from using equations (38) and (39) to relate the fluctuations in the density 
contrast to the fluctuations in $\psi$. Again, this is equivalent to the usual statement made about the power spectrum in terms of the reduced Mukhanov-Sasaki variable $\nu$ 
(namely that $ k^3 |\nu_{k}|^2  \sim \frac{ H^2 }{\epsilon_{SR} m_{pl}^2} $, for $m_{pl}$ Planck mass).
It is worthwhile to notice that the corresponding result for the tensor amplitudes will not be enhanced by a slow-roll factor, so we will ignore them in what follows. 

Using equation (55) we can average over the quadratic combinations of linear fluctuations to solve the constrained equation of motion for the diagonal second order 
homogeneous metric fluctuation ${\cal{Q}}_{0}(t)$, equation (48). In other words, our strategy is to use the first and second order constraints to substitute for the matter 
fluctuations which appear in the Einstein equation for ${\cal{Q}}_{0}(t)$, and then solve for ${\cal{Q}}_{0}(t)$ to calculate $\delta^2 \rho$and $\delta^2 p$

Indeed, keeping only the dominant terms and substituting for the matter terms, 
\begin{eqnarray}
(\partial^2_{t} + H \partial_{t} ) {\cal{Q}}_{0}(t) &=& 4 \pi \int_{k_{min}}^{aH} {}^{(2)}S(k) k^2 dk,    
\end{eqnarray}
where ${}^{(2)}S(k) = H^2  \left[ 24Ht -162\frac{ k^2 }{a^2 H^2 } \left( \frac{H^4}{\kappa \beta^2} \right) \right] |\psi_{k}|^2$ and $k_{min} << aH$. Carrying out the integral 
over $k$ and solving equation (56), it is relatively straightforward to show that the dominant solution takes the form
\begin{eqnarray}
{\cal{Q}}_{0}(t) &\approx& \frac{ \kappa H^2 \alpha}{4\pi \epsilon_{SR}^2} \left( \frac{ 27^2}{10} - 6 N \right) 
									- \frac{ 36 N \kappa H^2 }{4\pi \epsilon_{SR}^2 } \ln (\frac{aH}{k_{min}} ),  
\end{eqnarray}
where $N = \int H dt = \alpha \ln(t)$ is the number of e-foldings, and $\alpha = Ht >> 1$ as described in Section II. Using this result and performing similar integrations for 
the remaining terms in equations (50) and (51), we finally obtain expressions for the contributions to the homogeneous mode of $\delta^2 \rho$ and $\delta^2 p$
\begin{eqnarray}
\nonumber
\delta^{2} \rho_{IR} &\approx&     -\frac{3H}{\kappa} (H + \partial_{0}){\cal{Q}}_{0}(t) - \int_{k_{min}}^{aH} \frac{54 H^6 }{\kappa^2  \beta^2 }  4 \pi |\psi_{k}|^2 k^2 dk    \\
&\approx& \frac{H^2}{\kappa} \left( \frac{\kappa H^2 }{ 4 \pi \epsilon_{SR}^2 } \right) 
							\left[ -3 \alpha( \frac{27^2}{10} - 6N) + 36(N -\frac{27}{2}) \ln(\frac{aH}{k_{min}}) \right] \\
\nonumber	
\delta^{2} p_{IR} + \delta^{2} \rho_{IR} &\approx&   \int _{k_{min}}^{aH} 
				\frac{216 H^4 }{a(t)^2  \kappa^2 \beta^2 \epsilon_{SR}} ( 4 \pi k^4 |\psi_{k}|^2 )dk \\
&\approx& 2 \pi \frac{216 H^4 }{a^2  \kappa^2 \beta^2 \epsilon_{SR}}  (a H)^2 \left( \frac{9 \kappa H^2 }{16 \pi^2} \right), 
\end{eqnarray}
We can immediately  compare the magnitude of the leading homogeneous backreaction term in, say, $\delta^2 \rho_{IR} / \rhobar$, to the root mean square of the 
density contrast $\sqrt{ <\left( \delta \rho / \rhobar \right)^2> }$ during inflation by using equation (55).   One can demand a consistency condition 
for linearized theory, namely that the second order contributions be subdominant compared to that of the linearized sector, 
i.e. $\delta^2 \rho_{IR} \  < \   \ \ \sqrt{ <\left( \delta \rho \right)^2> }$. If we use the expressions (55), (57) and (58) above, 
this demand is crudely equivalent to the condition that, for $N \stackrel{>}{\sim} 70$, 
\begin{eqnarray}
									\epsilon_{SR}  > \    (4 \kappa H^2)^{1/4} N^{3/4},
\end{eqnarray}
i.e. the usual slow-roll condition will in general be violated if $\sim (\kappa H^2)^{1/4} N^{3/4} > 1$.  Thus 
inequality (60) suggests, within our scheme, that the breakdown of the linearized approximation occurs when one assumes the slow-roll condition for the background 
spacetime.
Furthermore it is apparent that, although the right hand side of (57) is not zero, the form of these dominant contributions is approximately that of a 
cosmological constant since $( \delta^2 p + \delta^2 \rho ) / \delta^2 \rho  \sim - 1 / \ln( \frac{k_{min}}{aH} ) \sim 0$.  

As we alluded to above, in the current gauge fixing we have chosen the homogeneous limit of the fluctuations is not well
defined since, by equation (55) and the linear order equation $\psi = -A$, the lapse $A$ will diverge as $k \rightarrow 0$. However, our results are 
valid even if one makes a gauge transformation which renders the superhorizon fluctuations well defined in the homogeneous limit. Indeed, after making such a gauge 
transformation,hen the homogeneous limit is taken and the total lapse goes to the value $1$ while the offdiagonal terms go to $0$, the central feature of `slow-roll` 
enhancement remains and the above arguments still apply. Let us consider how this works explicitly by looking at the linear case. Choosing the infinitesimal gauge 
transformations to be 
$\zeta^{a} = (T(t,k),k L(t,k),k L(t,k),k L(t,k))$, we first note that $T_{k}(t)$ must be, by equation (55) and the fact that 
$\delta g^{\prime}_{00} = \delta g_{00} - 2 \dot{T}$, so that
\begin{eqnarray}
T_{k}(t) &=& - \frac{ 3 \ln a}{4 \pi} \sqrt{\frac{\kappa}{k^3 \epsilon_{SR}}}
\end{eqnarray}
ensures that $|A_{k}| \rightarrow 0$ for $k \rightarrow 0$. Given this divergent gauge transformation of the comoving time, we can ask for what $L_{k}(t)$ we can can ensure 
the rest of the perturbations that appear in the metric will be well-defined for $k \rightarrow 0$. Since 
\begin{eqnarray}
\lie_{\zeta} \bar{g}_{0i} &=& -T_{,i} + a^2 \dot{L}_{,i}, 
\end{eqnarray}  
in position space, then choosing $L_{k}(t) = - \int \frac{\ln a} {k^{3/2}} \frac{ 3 }{4 a^2 \pi} \sqrt{\frac{\kappa}{\epsilon_{SR}}} dt$ will render the shift $B$ 
induced by equation (60) to zero. 
The off-diagonal terms $E$ induced in turn by this choice 
of $L$ will be well posed in the homogeneous limit since $k^2 E$ is what appears in the metric, and this will decay as $\sqrt{k}$. 
However, the diagonal spatial metric terms will in general receive a 
contribution of $2Ha^2 T$, which will get large as $k \rightarrow 0$. Since we implicitly use an IR cutoff of $k_{min}$, beyond which equation (54) will not hold and 
may in fact be replaced by a relation which does not diverge with $k$, these large contributions can be considered in some sense regulated. 

To second order the argument is very similar, only more tedious. Given the above choice for the linear gauge-fixing as $k \rightarrow 0$, we must pick a second order 
$\chi^{a} = ({}^{(2)}T(t,k),0,0,0)$ such that the second order shift and offdiagonal spatial terms go to zero in the limit $k \rightarrow 0$. Since 
\begin{eqnarray}
\nonumber
\delta^{2} \tilde{g}^{\prime}_{0i} &=& \delta^{2} g_{0i} +  \delta g_{00} k_{i} T + \delta g_{ii} k^{i} \dot{L} + \delta g_{ij}^{(i \neq j)} k^j \dot{L}
					+ \delta g_{i0} ( \dot{T} - 2 k^2 L) + T \partial_{0} \delta g_{0i} - k {}^{(2)} T + {}^{(2)} S
\end{eqnarray}
where ${}^{(2)} S \equiv  a^2 k \dot{L} (\dot{T} - 12 k^2 L) - kT(4 \dot{T} - 4a \dot{a} \dot{L} - 6k^2L - a^2 \ddot{L})$, one can show that the choice 
\begin{eqnarray}
{}^{(2)} T  = \frac{1}{k} \left[ {}^{(2)} S + Ak T + 3 \psi k \dot{L} \right]
\end{eqnarray}
will yield $\delta^2 \tilde{g}^{\prime}_{0i} = 0$.  The offdiagonal, second order, spatial terms induced by the above transformation are
\begin{eqnarray}
\delta^2  \tilde{g}^{\prime}_{ij} &\stackrel{i \neq j}{=}& 2k^2 T^2 - 4ak^2(2 \dot{a} L + a \dot{L})T + 18 a^2 k^4 L^2,
\end{eqnarray}
which at worst diverge as $k^{-1}$. Since only the expression $k^2( {}^{(2)} E )$ appears in the metric, these offdiagonal terms in the metric 
smoothly go to zero in the homogeneous limit. Once again the spatial diagonal contributions to ${\cal{Q}}$ will grow large as $k \rightarrow 0$. 

Now we can finally address the total effect of all these transformations on the quantities of interest, $\delta^2 \rho, \delta^2 p$. To second order, for example, 
$\delta^2 \rho$ will be Lie-dragged along $\zeta^{a}$ and $\chi^{a}$ according to the tranformation  
\begin{eqnarray}
\delta^2 \tilde{\rho}^{\prime} &=& \delta^2 \rho + ( \lie^2_{\zeta} + \lie_{\chi} ) \rhobar + 2 \lie_{\zeta} \delta \rho \\
\nonumber
&=& \delta^2 \rho + 2( T \partial_{0} \delta \rho - 3 k^2 L \delta \rho ) + {}^{(2)} T \dot{\rhobar} 
				+ \left( T^2 \ddot{\rhobar} + T \dot{T} \dot{\rhobar} - \dot{\rhobar} k^2 LT \right)
\end{eqnarray}
and similarly for the averaged pressures $\delta^2 p$. Choosing the above expressions for $T, L, {}^{(2)}T$, we can see that the total effect on $\delta^2 \rho$ and $\delta^2 p$
will be to introduce terms that diverge like $k$ but are suppressed by factors of $\beta$. Such terms will be subdominant compared to the terms already present in the original
gauge, or in other words, the gauge transformation which renders the metric diagonal in the homogeneous limit does not undo the dominant contributions to the energy density and 
pressure at second order. 

\subsection{ Comment on backreaction on inhomogeneous second order modes }

An important loophole in the above analysis resides in the fact that we compare the horizon scale (frozen) amplitudes of the linearized fluctuations to that of the 
homogeneous sector of the second order fluctuations. It is far from clear that this is an acceptable comparison to make, not least because we are directly 
cutting off the divergence of the linearized fluctuations by imposing an IR cutoff at $k = k_{min} << aH$ and then comparing the amplitude of the second order fluctuations 
with that of the linearized fluctuations well beyond the cutoff. A valid criticism of this result is thus that it would be natural for the second order 
perturbations to dominate at the homogeneous scale simply because we have cut off the linearized fluctuations long before comparing their amplitude to that at 
second order, or in other words the spatial dependence induced by evaluating the (quantum averaged) second order amplitude at some $k_{min} < k = \kt << aH$ may 
alter the conclusions of inequality (60).   See Figure 1. 

However, we find that when we indeed compare the quantity $\sqrt{< \left( \frac{ \delta^2 \rho_{IR} }{\rhobar} \right)^2 >}$ at some scale 
$\kt$ such that $k_{min} \stackrel{<}{\sim} \kt << aH$ to the horizon-scale amplitude of the linearized fluctuations   
$\sqrt{< \left . \left( \frac{ \delta \rho }{\rhobar} \right)^2 \right|_{k=aH} >}$, we once again find that the amplitude of the second order modes may still dominate 
over that of the linear modes assuming slow-roll.
\begin{figure}
\epsfxsize=45mm
\epsfysize=45mm
\epsfbox{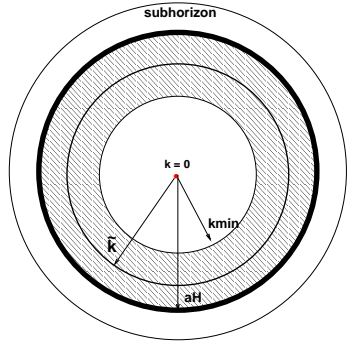}
\caption{ All linearized modes in the shaded region, spanning from $k=aH$ to $k=k_{min} << aH$ in spatial scale, are taken to seed second order modes at a particular 
value of $k$. The thick black boundary indicates the Hubble scale, where the amplitudes of the linearized fluctuations freeze out during slow-roll. 
We compare the amplitudes of the linearized fluctuations at the thick black boundary to that of the second order fluctuations at $k=\kt$ and $k=0$ (and ignore the 
influence of suitably renormalized subhorizon ($k> aH$) modes at second order) during slow-roll.    }
\label{fig:test}
\end{figure}
As show in detail in the Appendices, we find that we may write 
\begin{eqnarray}
\nonumber
< \delta^2 p_{IR}(k) \delta^2 p^{\dagger}_{IR}(k) > &\approx& 
			< \intkp \intkpp \left( \frac{ 54 H^2}{\kappa \epsilon_{SR}} \right)^2 \psikpminusk \psikp \psikppminusk \psikpp d^3 \vec{\kp} d^3 \vec{\kpp} > \\
\nonumber
&+& <\intkp \left( \frac{ 3H}{\kappa} \right)^2 {\cal{L}} {\cal{Q}}_{\kp - k}  {\cal{L}} {\cal{Q}}_{\kp} d^3 \vec{\kp}  > \\
&+& 
< \intkp \intkpp \left( \frac{ 6H}{\kappa}  \frac{ 54 H^2}{\kappa \epsilon_{SR}} \right) {\cal{L}} {\cal{Q}}(\tp, \kp;k) \psikppminusk \psikpp d^3 \vec{\kp} d^3 \vec{\kpp}   >,
\end{eqnarray}
where ${\cal{L}} \equiv \left( \partial_{0} + H \right)$, and that this in turn has the form 
\begin{eqnarray}
< \delta^2 p_{IR}(k) \delta^2 p^{\dagger}_{IR}(k) > &\approx& 
\left( \frac{H^2}{\kappa} \right)^2 \frac{\kappa^2 H^4}{\epsilon_{SR}^4 \pi^2} \left( A_{1} \alpha^2 + \alpha \left( B_{1} \ln (\gamma) + C_{1} \ln (\sigma) \right) \right), 
\end{eqnarray}
where 
\begin{eqnarray}
A_{1} &\equiv& \frac{2657205}{1156 \pi^2} \approx 233 \\
B_{1} &\equiv& \frac{-1594323}{68 \pi^2 } \approx -2376 \\
C_{1} &\equiv& \frac{2657205}{68 \pi^2} \approx 3959, 
\end{eqnarray}
and where 
\begin{eqnarray}
\sigma &\equiv& \frac{aH}{k_{min}} \\
\gamma &\equiv& 1 + \frac{2k}{k_{min}} 
\end{eqnarray}
Using all the above we can now finally see that the demand that 
$\sqrt{< \left . \left( \frac{ \delta^2 \rho_{IR} }{\rhobar} \right)^2 \right|_{k = \kt}} <  \sqrt{< \left . \left( \frac{ \delta \rho }{\rhobar} \right)^2 \right|_{k=aH} >}$, 
as described in Figure 1, is equivalent to 
\begin{eqnarray}
\epsilon_{SR} > \frac{2}{3} ( \kappa H^2 )^{\frac{1}{4}} {(A_{1}N) }^{\frac{1}{4}}
\end{eqnarray}
Given that $N \sim 70, A_{1} \sim 200$, (73) implies e.g. that for $\kappa H^2 \sim 1 \leftrightarrow H^2 \sim {m_{planck}}^2$ the slow-roll condition must be violated.
We see that the amplitude of the second order flucuations in our quantity $\delta^2 p_{IR}$  
dominates the corresponding linearized amplitude if the background spacetime is rolling slowly enough and $\kappa H^2$ is large enough, as it is in many models.  Note 
also the appearanec of the number of e-foldings $N$, which indicates, as previously shown, that this effect is a cumulative effect which depends on the growth of the 
phase space of superhorizon modes. All in all, the violation of this inequality in such models casts doubt on the viability of the linearized approximation to those slowly 
rolling spacetimes. 

As we imply in the caption to Figure 1 above, this calculation would ultimately contain formally divergent subhorizon contributions to $\delta^2 \rho$, $\delta^2 p$ 
and it is this real phsyical effect of the coupling between 
subhorizon (UV) modes with superHubble (IR) modes that will reveal the observable importance of backreaction effects for local observers. The points we are making here are  
1) the superHubble 
contribution, in this reasonable gauge, does take the approximate form of a cosmological constant contribution; and 2) assuming the slow-roll condition in this gauge,
within a wide class of initial conditions, is 
equivalent to assuming that IR backreactions dominate the linear terms. Even if one uses the 
residual homogeneous gauge freedom, one cannot simulataneously gauge away $\delta^2 \rho$ and $\delta^2 p$. These results hold even if one considers backreaction onto
some spatial scale $\kt << k_{min}$ of the second order modes.

\section{ Summary and Conclusion }

To summarize, we study the perturbations of the inflationary slow-roll spacetime
which are at second order in the metric and matter fluctuations. We follow a procedure of consistently (though probably not convergently) 
expanding the Einstein equations to second order and solving a subset of them assuming the zeroth and linear order equations hold. Specifically, we solve the 
first and second order Einstein constraint equations for the matter fluctuations, and in that way express our final expressions of $\delta^2 \rho$ and $\delta^2 p$ 
in terms of the metric fluctuations. We then solve a second order Einstein to express all of our expressions in terms of quadratic combinations of linearized metric 
fluctuations.

In order to isolate the physical degrees of freedom in the second order fluctuations we use a longitudinal gauge-fixing procedure at second order.  
Namely, we specify {\it two} independent infinitesimal, inhomogeneous,  coordinate transformations (gauges), one at linear order and one at 
second order, which admit no residual coordinate freedoms. Within this coordinate system we evaluate the fluctuations of two independent 
background scalars formed from the stress energy and from them define the fluctuations, to second order, of the isotropic energy density
$\delta^{2} \rho$ and pressure $\delta^2 p$. These fluctuations will not only arise from second order scalar modes 
but also from quadratic combinations of scalar-scalar and tensor-tensor modes at second order. Whereas the nondecaying linear scalar 
and tensor modes are constants at these scales (the vector modes die away), the nondecaying second order modes are time dependent and this 
leads to time-dependent $\delta^2 \rho$ and $\delta^2 p$. We calculate $\delta^2 \rho$ and $\delta^2 p$ using the linearized and second order constraints and one second order 
scalar Einstein equation. 
  
Futhermore, given that we have three effective small parameters in this problem (a slow-roll parameter, the strength of the metric and matter fluctuations, and the 
longwavelength approximation ($H^2 / \lambda^2 << 1$), we find that our $\delta^2 \rho$ and $\delta^2 p$  depend 
sensitively on the hierarchy of small parameters one assumes in the sense that ambiguous terms like 
$( k^2 / H^2 a^2 )(H^4 / \kappa \beta^2)$ appear in the mode expansion of $\delta^{2} \rho$, $\delta^{2} p$. For the (incomplete) case of just 
scalar-scalar backreactions and no genuinely second order metric or matter fluctuations, we find that $\delta^2 \rho < 0$ but that $\delta^2 p + \delta^2 \rho \neq 0$ in 
general. We find that the second order contributions to the energy density and pressure can, with the assumption of slow-roll, dominate over the 
second order linear contributions to the energy density and isotropic pressure given a broad range of slow-roll parameters.  It is worth emphasizing that the divergence of  
the enegy density or pressure perturbations as the slow-roll parameter goes to zero is certainly not physical, as our expressions (say, equation (67)) would seem to suggest. 
Rather, if one thinks of a single set of Einstein constraints expanded order by order (as opposed to a tower of constraints) then it may well be that second order terms in these 
constraints simply start to dominate the linear ones.

We conclude that when one truly goes to second order and solves some of the Einstein equations for the higher order classical fluctuations, 
they do approximately lead to a cosmological constant type of contribution in this gauge. Furthermore, it seems that these higher order corrections dominate the linear 
terms if slow-roll holds in the background, suggesting the breakdown of the linearized approximation to within a small window of slow-roll parameters. Some previous 
calculations of higer order superhorizon effects
(of which we are aware) have used a procedure which effectively takes the expectation value of the gauge-fixed metric 
{\it before} forming some sort of `invariant` measure of the expansion with which to probe 'local' backreaction, i.e. gauge fixing before taking the expectation value. 
Such a procedure suffers from higher order gauge ambiguities, and at least for a model with massless, minimally coupled scalar with quartic self-interaction 
(no gravity) one can first form a desired 'invariant' and then take expectation values and gauge fix this result (see for example \cite{Abramo:2001dd}, and also 
\cite{Geshnizjani:2002wp} ). Further investigations along this line will almost certainly prove useful, and one pay-off seems to be new ideas for observables in backreactions, 
such as recently described in \cite{Tsamis:2005bh}. 

Finally, as we indicated in the Introduction, the procedure we use here does not say anything
about what an observer would measure as the averaged cosmological constant in his own obserable patch of the universe. In other words, the suitably renormalized 
subhorizon contributions to $\delta^2 p$ and $\delta^2 \rho$ will allow a probe of the 
possibly crucial physics of the coupling of subhorizon and superhorizon modes in inflation and possibly shed some light on what we even mean by local modifications 
to a cosmological constant.

\section{ Acknowledgements }
Both authors thank NSERC and UBC for financial support during the course of this work. W.G.U. thanks the CIAR program for support. 
We also to thank R. Brandenberger, R. Woodard, and A.O. Barvinsky for many useful discussions and constant encouragement.

\appendix

\section{Four-point correlation functions}

The central result of this paper lies in comparing the second order quantity $\sqrt{< \left( \frac{ \delta^2 \rho_{IR} }{\rhobar} \right)^2 >}$ at some scale $\kt$ such that 
$k_{min} < \kt << aH$ to the horizon-scale amplitude of the linearized fluctuations   
$\sqrt{< \left . \left( \frac{ \delta \rho }{\rhobar} \right)^2 \right|_{k=aH} >}$. The angled 
brackets indicates averaging over a Hadamard vacuum state $|0>$. In this Appendix we supply details of the calculation of the four-point function which are 
inherent in the expressions for the dominant terms of $< \left( \frac{ \delta^2 \rho_{IR} }{\rhobar} \right)^2 >$. We first derive the expressions for the inhomogeneous 
case ($k \neq 0$) and then we take the homogeneous limit ($k = 0$). The Appendix ends with a section which proves that second order scalar metric perturbations are non-trivial
in pure de-Sitter (i.e., in a no-roll background), contrary to some claims made in the literature.

The case of the 2-point function is relatively simple to consider. Expanding $\psi_{k}$ in terms of the creation and annihilation operators $a, \ad$ we write 
\begin{eqnarray}
\psi_{k} = \omegak \ak + \omegastarkp \adkp,
\end{eqnarray}
where $\omega_{k}$ is equal to ( by analogy to the familiar Minkowski result $\omega_{k} \sim \left(  \frac{e^{ikx}}{\sqrt{2k}} \right)$) 
\begin{eqnarray}
\nonumber
\omega_{k} = \sqrt{\frac{1}{a^{3}}} \sqrt{ \frac{a}{2k}} e^{ik/aH} \left(1 + \frac{iaH}{k} \right)
\end{eqnarray}
The first term is a normalization term that goes as $1 / \sqrt{V}$ since $a^{3}$ corresponds to the volume measure of a comoving observer. This solution to equation (37) 
(with the spatial gradient term restored) is valid to within a couple of Hubble times of the horizon exit, and during this time the variation of $H$ is negligible.
Therefore it makes sense that, {\it up to some phase factor that varies slowly on the Hubble timescale}, the expression of $\omega_{k}$ has this simple form
compared to the flat space result. 

Defining the 2-point function as 
\begin{eqnarray}
\nonumber
<\psi^2> \equiv < \intkp \psikpminusk \psikp d^3 \vec{\kp} >  
\end{eqnarray}
we find that 
\begin{eqnarray}
< \psi^2> &=&  < \intkp (\akpminusk \omegakpminusk + \adkpminusk \omegastarkpminusk)(\akp \omegakp + \omegastarkp \adkp) d^3 \vec{\kp} > \\
\nonumber
&=& < \intkp \akpminusk \adkp \omegakpminusk \omegastarkp d^3 \vec{\kp} > \\
\nonumber
&=& \delta( -k ) < \intkp \omegakpminusk \omegastarkp d^3 \vec{\kp} >, 
\end{eqnarray}
where $\Omega_{\kp}$ indicates the superHubble range of integration for $\kp$, namely $\kp \in [k_{min}+k, aH], k_{min} << aH$. In other words, 
the quantum average of the 2-point function is only non-trivial for homogeneous contributions, as suggested by the Poincare invariance of the linearized fluctuations. In the 
following we will assume that the spectrum of modes is discrete by imposing periodic boundary conditions ($x \rightarrow x + L$) on the spatially flat slicing of the 
background. In this manner the delta functions that appear in the commutation relations of ladder operators simply become Kronecker deltas and the volume normalizations are 
implicitly carried by the above definition of $\omega_{k}$. 

For the four-point function the situation is different. More specifically, we wish to compute the `square of the two-point function`, i.e. we do not want technically want the 
four-point function but rather the {\it fluctuations} in the operator $\psi^{2}$ since we want to measure and compare the {\it dispersions} of $\rho, p$ at first and 
second order. Indeed, denoting these fluctuations by $<\psi^4>$ we define (in the continuum limit)
\begin{eqnarray}
\nonumber
< \psi^4 > &\equiv& < \intkp \intkpp \psikpminusk \psikp \psikppminusk \psikpp d^3 \vec{\kp} d^3 \vec{\kpp} >,   
\end{eqnarray}
one may verify that 
\begin{eqnarray}
< \psi^4 > &=& \left( < \intkp \psikpminusk \psikp d^3 \vec{\kp} > \right)^2  \\
\nonumber
&+& < \intkp \intkpp \psikpminusk \psikppminusk d^3 \vec{\kp} d^3 \vec{\kpp} >< \intkp \intkpp \psikp \psikpp d^3 \vec{\kp} d^3 \vec{\kpp} > \\
\nonumber
&+& <\intkp \intkpp \psikpminusk \psikpp d^3 \vec{\kp} d^3 \vec{\kpp} >< \intkp \intkpp \psikppminusk \psikp d^3 \vec{\kp} d^3 \vec{\kpp} >, 
\end{eqnarray}
where similarly $\Omega_{\kpp}$ indicates the superHubble range of integration for $\kpp$, namely $\kpp \in [k_{min}+k,aH]$. Assuming that $k \neq 0$ we find that, using 
equation  (A2), 
\begin{eqnarray}
\nonumber
< \psi^4  > &=& <\intkp \intkpp \akp \adkpp \omegakp \omegastarkpp d^3 \vec{\kp} d^3 \vec{\kpp} >
										< \intkp \intkpp \akpminusk \adkppminusk \omegakpminusk \omegastarkppminusk d^3 \vec{\kp} d^3 \vec{\kpp} > \\
\nonumber
&+& <\intkp \intkpp \akpminusk \adkpp \omegakpminusk \omegastarkpp d^3 \vec{\kp} d^3 \vec{\kpp} >< \intkp \intkpp \akppminusk \adkp \omegakppminusk \omegastarkp d^3 \vec{\kp} d^3 \vec{\kpp} >,
\end{eqnarray}
which straightforwardly simplifies to 
\begin{eqnarray}
< \psi^4 > &=& \left( \intkpp |\omegakppminusk|^2 d^3 \vec{\kpp} \right) \left[ \frac{1}{4} \frac{9\kappa}{\epsilon_{SR}} \left( \frac{H}{2\pi} \right)^2 4 \pi
\ln(\frac{aH}{k_{min}}) + \intkp |\omegakpminusk|^2 d^3 \vec{\kp} \right] 
\end{eqnarray}
using equation (55) in the main text for $k > k_{min}$. It turns out that we shall also require the expressions $< \psikpminusk \psikp \psikppminusk \psikpp {\kp}^2>$ and 
$<\psikpminusk \psikp \psikppminusk \psikpp {\kpp}^2 {\kp}^2>$ in what follows, so, using the above, we find (for $k \neq 0$)
\begin{eqnarray}
\nonumber
< \psikpminusk \psikp \psikppminusk \psikpp {\kp}^2> &=&  \left( \intkp {\kp}^2 |\omegakpminusk|^2 d^3 \vec{\kp} \right) \left( \intkpp |\omegakppminusk|^2 d^3 \vec{\kpp} \right) \\
\nonumber
&&
+  \left( \intkp  |\omegakpminusk|^2 d^3 \vec{\kp} \right) \left( \intkpp \kpp^2 |\omegakppminusk|^2 d^3 \vec{\kpp} \right) \\
\nonumber
&&
+  \left( \intkp  \kp^2 |\omegakp|^2 d^3 \vec{\kp} \right) \left( \intkpp  |\omegakppminusk|^2 d^3 \vec{\kpp} \right) \\
&&
+  \left( \intkp   |\omegakp|^2 d^3 \vec{\kp} \right) \left( \intkpp  \kpp^2 |\omegakppminusk|^2 d^3 \vec{\kpp} \right) 
\end{eqnarray}
and
\begin{eqnarray}
\nonumber 
<\psikpminusk \psikp \psikppminusk \psikpp {\kpp}^2 {\kp}^2> &=& \left( \intkp \kp^2 |\omegakpminusk|^2 d^3 \vec{\kp} \right) \left( \intkpp \kpp^2 |\omegakpp|^2 d^3 \vec{\kpp} \right) \\
\nonumber
&&
+ \left( \intkp \kp^4 |\omegakpminusk|^2 d^3 \vec{\kp} \right) \left( \intkpp |\omegakpp|^2 d^3 \vec{\kpp} \right) \\
\nonumber
&&
+ \left( \intkp  |\omegakpminusk|^2 d^3 \vec{\kp} \right) \left( \intkpp \kpp^4 |\omegakpp|^2 d^3 \vec{\kpp} \right) \\
\nonumber
&&
+ \left( \intkp \kp^2 |\omegakpminusk|^2 d^3 \vec{\kp} \right) \left( \intkpp \kpp^2 |\omegakppminusk|^2 d^3 \vec{\kpp} \right) \\
\nonumber
&&
+ \left( \intkp \kp^4 |\omegakpminusk|^2 d^3 \vec{\kp} \right) \left( \intkpp |\omegakppminusk|^2 d^3 \vec{\kpp} \right) \\
&&
+  \left( \intkp  |\omegakpminusk|^2 d^3 \vec{\kp} \right) \left( \intkpp \kpp^4 |\omegakppminusk|^2 d^3 \vec{\kpp} \right)
\end{eqnarray}

Once again taking equation (55) to give the amplitude of the linearized quantum fluctuations at horizon crossing and assuming this frozen amplitude for $\kp, \kpp << aH$ 
(up until $\kp, \kpp = k_{min}$, where we cut it off) we complete the calculation, using  
\begin{eqnarray}
\nonumber
\intkp |\omegakpminusk|^2 d^3 \vec{\kp} &=& 
\nonumber
   \intkp \left[ \int_{0}^{\pi} \int_{0}^{2 \pi} \frac{1}{ \left( k^2 + \kp^2 -2k \kp \cos(\theta) \right)^{3/2} } d \phi \sin( \theta ) d \theta  \right] \kp^2 d \kp \\
\nonumber
&=& \frac{1}{4} \frac{9\kappa}{\epsilon_{SR}} 
			\left( \frac{H}{2\pi} \right)^2 4 \pi \left. \left( \frac{1}{2} ( \ln(\kp - k) + \ln(\kp + k) ) \right) \right|^{aH}_{k_{min}+k} \\
&\approx& \frac{1}{4} \frac{9\kappa}{\epsilon_{SR}} 
			\left( \frac{H}{2\pi} \right)^2 4 \pi  \left( \ln \frac{aH}{k_{min}} - \frac{1}{2} \ln (1 + 2 \frac{k}{k_{min}}) \right), 
\end{eqnarray}
and similarly 
\begin{eqnarray}
\intkp \kp^2 |\omegakpminusk|^2 d^3 \vec{\kp} &=& \frac{1}{4} \frac{9\kappa}{\epsilon_{SR}} \left( \frac{H}{2\pi} \right)^2 2 \pi
			\left . \left( \kp^2 + k^2 ( \ln(\kp - k) + \ln(\kp + k) )  \right) \right|_{k_{min}+k}^{aH}  \\
\nonumber
&\approx&  \frac{1}{4} \frac{9\kappa}{\epsilon_{SR}} \left( \frac{H}{2\pi} \right)^2 2 \pi 
			\left( (aH)^2 + 2 k^2  \left( \ln \frac{aH}{k_{min}} - \frac{1}{2} \ln (1 + 2 \frac{k}{k_{min}}) \right)  \right), 
\end{eqnarray}
along with (remembering again that $k << aH$)
\begin{eqnarray}
\nonumber
\intkp \kp^4 |\omegakpminusk|^2 d^3 \vec{\kp} &=& \frac{1}{4} \frac{9\kappa}{\epsilon_{SR}} \left( \frac{H}{2\pi} \right)^2 \pi
		\left . \left( \kp^4 + 2 \kp^2 k^2 + 2k^4  ( \ln(\kp - k) + \ln(\kp + k) )  \right) \right|_{k_{min}+k}^{aH} \\
&\approx&  \frac{1}{4} \frac{9\kappa}{\epsilon_{SR}} \left( \frac{H}{2\pi} \right)^2 \pi 
			\left( (aH)^4 + 4 k^4  \left( \ln \frac{aH}{k_{min}} - \frac{1}{2} \ln (1 + 2 \frac{k}{k_{min}}) \right) \right)
\end{eqnarray}
to obtain finally (using the fact that the $\kp, \kpp$ ranges of integration are identical)
\begin{small}
\begin{eqnarray}
< \psikpminusk \psikp \psikppminusk \psikpp> &\approx& \eta \left( 2 (\ln(\sigma))^2 - \frac{3}{2} \ln(\sigma) \ln(\gamma) + \frac{1}{4} \left( \ln(\gamma) \right)^2 \right) \\
%\nonumber
< \psikpminusk \psikp \psikppminusk \psikpp \kp^2> &\approx& \eta (aH)^2 \left( \left[1 + 2 \left(\frac{k}{aH}\right)^2 \ln \left(\frac{\sigma}{\sqrt{\gamma}}\right) \right]
					\ln  \left( \frac{\sigma^2}{2\sqrt{\gamma}}\right)  + \frac{1}{2} \ln \left( \frac{\sigma}{\sqrt{\gamma}} \right) \right) \\
\nonumber
< \psikpminusk \psikp \psikppminusk \psikpp {\kp}^2 \kpp^2> &\approx& \eta (aH)^4 
\left[ \left( \frac{1}{2} + \frac{1}{4} ( \ln (\sigma) + 3 \ln \frac{\sigma}{\sqrt{\gamma}} ) \right) 
		+ \left(\frac{k}{aH}\right)^2 \left( \frac{3}{2} \ln \frac{\sigma}{\sqrt{\gamma}} \right) \right . \\
&& \ \ \ \ \ \ \ \ 
\ \ \ \ \ \left. +  \left(\frac{k}{aH}\right)^4 \left( 3  \left( \ln \frac{\sigma}{\sqrt{\gamma}} \right)^2 +  \ln \sigma \ln \frac{\sigma}{\sqrt{\gamma}} \right) \right]
\end{eqnarray}
\end{small}
In the above, we define the dimensionless factors $\sigma, \gamma, \eta$ as
\begin{eqnarray}
\sigma &\equiv& \frac{aH}{k_{min}} \\
\gamma &\equiv& 1 + \frac{2k}{k_{min}} \\
\eta &\equiv& \left( \frac{9 \kappa \pi}{\epsilon_{SR}} \right)^2 \left( \frac{H}{2 \pi} \right)^4, 
\end{eqnarray}
and once again we assume $k_{min} << aH$. We retain powers of $k / a H$ for now simply for generality. The long-wavelength approximation will kill off these terms later on in 
the calculation.  

Finally, one can take the homogeneous limit ($k \rightarrow 0$) of expressions (A9)-(A11), bearing in mind that the squares of the two-point 
functions now contribute as shown by equation (A2). When we take the homogeneous limit of the above equations we find 
\begin{eqnarray}
< \psi^{4} >_{k = 0} &\approx& 3 \eta \left( \ln\sigma \right)^2 \\
\lim_{k \rightarrow 0} < \psikpminusk \psikp \psikppminusk \psikpp \kp^2> &\approx& 3 \eta (aH)^2 \ln \sigma \\
\lim_{k \rightarrow 0} < \psikpminusk \psikp \psikppminusk \psikpp \kp^2 \kpp^2> &\approx& \frac{3}{2} \eta (aH)^4 \left( \frac{1}{2} + \ln \sigma \right)
\end{eqnarray}
which provides a coarse but useful check on the algebra to this stage.

\section{Second order equations of motion and correlation functions at second order }

In order to compute the full quantity $\delta^2 \rho(k) \delta^2 \rho^{\dagger}(k)$ we shall formally encounter two-point functions not only involving $\psi_{k}$, but also 
those involving the second order fluctuations ${\cal{Q}}_{k}$, e.g. 
\begin{eqnarray}
< \intkp  {\cal{L}} {\cal{Q}}_{\kp - k}  {\cal{L}} {\cal{Q}}_{\kp} d^3 \vec{\kp} >,
\end{eqnarray}
where ${\cal{L}} \equiv \left( \partial_{0} + H \right)$. Of course, by solving the second order equations of motion these sorts of expressions can be 
reduced to four-point functions involving only $\psi_{k}$. We now show how this reduction is accomplished.

Using equation (56) and the fact that $\dot{H} = -H^2 \frac{\epsilon_{SR}}{18}$, one can easily show via by-parts integration\footnote{ In operating on expressions involving 
$H, a$, etc., we always first assume the time dependence of $a(t) = a_{0} t{\alpha}, H(t) = \alpha / t$ before the operation, and only after the operation take the limit of 
$\alpha >> 1, \alpha / t \rightarrow H \in \Re$. So, for example, $\int H dt = \alpha \ln(t) \neq H t$ {\it before} taking the limit $\alpha >> 1$. } that 
\begin{eqnarray}
{\cal{L}} {\cal{Q}}_{k} &=& \int^{t} \intkp {}^{(2)} S(\tp, \kp;k) d^3 \vec{\kp} d \tp 
						+ \frac{1}{18} \int^{t} \intkp \epsilon_{SR} H(\tp)^2 {\cal{Q}}(\tp,\kp;k) d \tp d^3 \vec{\kp},
\end{eqnarray}
i.e. we compute the first integral of the reduced second order equation of motion.
A tedious integration reveals that the leading terms of the latter integral over ${\cal{Q}}$ are suppressed by a factor of $\epsilon_{SR} / \alpha$ compared to those of the 
first term. Therefore we ignore the latter terms and write 
\begin{eqnarray}
{\cal{L}} {\cal{Q}}_{k} &\approx& \int^{t} \intkp {}^{(2)} S(\tp, \kp;k) d^3 \vec{\kp} d \tp 
\end{eqnarray}
which immediately leads us to the expression, again using equation (56),
\begin{eqnarray}
< {\cal{L}} {\cal{Q}}_{k} > &=& 
     < \int^{t} \intkp H(\tp)^2 \left( 24  H(\tp) \tp - 162 \frac{{\kp}^2}{a(\tp)^2 H(\tp)^2} \frac{ H(\tp)^4}{\kappa \beta^2} \right) \psikpminusk \psikp d^3 \vec{\kp} d \tp > 
\end{eqnarray}
Equation (B4) allows us to evaluate the relevant quantities which will appear in the expression for $\delta^2 \rho(k) \delta^2 \rho(k)$, such as (B1) and 
$<\intkp \intkpp {\cal{L}} {\cal{Q}}(\tp, \kp;k) \psikppminusk \psikpp d^3 \vec{\kp} d^3 \vec{\kpp}>$. Notice that ${\cal{L}} {\cal{Q}}$ has units $1 / s$, as makes sense since 
the metric fluctuations are defined as dimensionless. Using the early results of Appendix A, we find that 
\begin{eqnarray}
\!\!\!\!\!\!\!\!\!\!\!\! < \intkp  {\cal{L}} {\cal{Q}}_{\kp - k} {\cal{L}} {\cal{Q}}_{\kp}  d^3 \vec{\kp} > &=&   
< \int^{t} H^2 \left( 24  H \tp - 162 \frac{{\kp}^2}{a^2 H^2} \frac{ H^4}{\kappa \beta^2} \right)  \intkp \psikpminusk \psikp d^3 \vec{\kp} d \tp \\
\nonumber
&&  
   \ \ \ \ \ \times \int^{t} H^2 \left( 24  H \tpp - 162 \frac{{\kpp}^2}{a^2 H^2} \frac{ H^4}{\kappa \beta^2} \right) \intkpp \psikppminusk \psikpp d^3 \vec{\kpp} d \tpp > 
\end{eqnarray}
\begin{eqnarray}
\nonumber
<\intkp \intkpp {\cal{L}} {\cal{Q}}(\tp, \kp;k) \psikppminusk \psikpp d^3 \vec{\kp} d^3 \vec{\kpp}> &=& 
< \int^{t} H^2 \left( 24  H \tp - 162 \frac{{\kp}^2}{a^2 H^2} \frac{ H^4}{\kappa \beta^2} \right)  \intkp \psikpminusk \psikp d^3 \vec{\kp} d \tp \\
&&
   \ \ \ \ \ \ \ \ \ \ \ \ \ \ \ \ \times \intkpp \psikppminusk \psikpp d^3 \vec{\kpp} > ,
\end{eqnarray}
where the appropriate time dependences of $a(t) \sim a_{0} t^{\alpha}, H(t) \sim \alpha / t$ are assumed above (as properly shown in equation (B4)). Expanding the above 
expressions (assuming once again that the ranges of integration for $\kp, \kpp$ are the same, as in Appendix A), we obtain 
\begin{eqnarray}
< \intkp  {\cal{L}} {\cal{Q}}_{\kp - k}  {\cal{L}} {\cal{Q}}_{\kp} d^3 \vec{\kp} > &=& 
\int^{t} \int^{t} H(\tp)^3 H(\tpp)^3 (24)^2 \tp \tpp  < \psi^4 > d \tp d\tpp \\
\nonumber
&& -  \int^{t} \int^{t} H(\tp)^4 H(\tpp)^3 (24 \tp) \left( \frac{162}{a(\tp)^2 \kappa \beta^2}  \right)  
												< \psi^4  \kp^2 > d \tp d \tpp\\
\nonumber
&& -  \int^{t} \int^{t} H(\tpp)^4 H(\tp)^3 (24 \tpp) \left( \frac{162}{a(\tpp)^2 \kappa \beta^2}  \right)  
												< \psi^4 \kpp^2 > d \tp d \tpp\\
\nonumber
&& + \int^{t} \int^{t} (162)^2 H(\tp)^4 H(\tpp)^4 \left( \frac{1}{a(\tp)^2 a(\tpp)^2 \kappa^2 \beta^4}  \right)  
												< \psi^4 \kp^2 \kpp^2 > d \tp d \tpp 
\end{eqnarray}
and similarly
\begin{eqnarray}
<\intkp \intkpp {\cal{L}} {\cal{Q}}(\tp, \kp;k) \psikppminusk \psikpp d^3 \vec{\kp} d^3 \vec{\kpp}> &=& 
						24 \int^{t} H^2 (H \tp) <\psi^4 > d \tp \\
\nonumber
&& - 162 \int^{t} H^2 \frac{1}{a^2 H^2} \frac{H^4}{\kappa \beta^2} < \psi^4 \kp^2 > d \tp ,
\end{eqnarray}
where we define $<\psi^4 > \equiv< \psikpminusk \psikp \psikppminusk \psikpp >, < \psi^4 \kp^2> \equiv < \psikpminusk \psikp \psikppminusk \psikpp \kp^2 >$ and 
$<\psi^4 \kpp^2 \kpp^2> \equiv  < \psikpminusk \psikp \psikppminusk \psikpp \kp^2 \kpp^2 >$ all integrated over $\kp, \kpp$, as in the above sections. 

Inserting expressions (A10) through (A12) (and using (A5) and (A6) to perform the temporal product integrations in (B7) [which amount to symmetrization in 
$\tp$, $\tpp$] ) into (B7) and (B8) we obtain (for $k \neq 0$)
\begin{small}
\begin{eqnarray}
&& < \intkp  {\cal{L}} {\cal{Q}}_{\kp - k} {\cal{L}} {\cal{Q}}_{\kp} d^3 \vec{\kp} > = 
\nonumber
\int^{t} \int^{t} H(\tp)^3 H(\tpp)^3 (24)^2 \tp \tpp  
				\tilde{\eta} \left( 2 (\ln(\tilde{\sigma}))^2 - \frac{3}{2} \ln(\tilde{\sigma}) \ln(\gamma) 
										+ \frac{1}{4} \left( \ln(\gamma) \right)^2 \right) d \tp d\tpp \\
\nonumber
&& -  \int^{t} \int^{t} H(\tp)^4 H(\tpp)^3 (24 \tp) \left( \frac{162}{a(\tp)^2 \kappa \beta^2}  \right)  
			\tilde{\eta} (\tilde{aH})^2 \left( \left[1 + 2 \left(\frac{k}{\tilde{aH}}\right)^2 \ln \left(\frac{\tilde{\sigma}}{\sqrt{\gamma}}\right) \right]
		\ln  \left( \frac{{\tilde{\sigma}}^2}{2\sqrt{\gamma}}\right)  + \frac{1}{2} \ln \left( \frac{\tilde{\sigma}}{\sqrt{\gamma}} \right) \right)d \tp d \tpp\\
\nonumber
&& -  \int^{t} \int^{t} H(\tpp)^4 H(\tp)^3 (24 \tpp) \left( \frac{162}{a(\tpp)^2 \kappa \beta^2}  \right) 
           \tilde{\eta} (\tilde{aH})^2 \left( \left[1 + 2 \left(\frac{k}{\tilde{aH}}\right)^2 \ln \left(\frac{\tilde{\sigma}}{\sqrt{\gamma}}\right) \right]
	\ln  \left( \frac{{\tilde{\sigma}}^2}{2\sqrt{\gamma}}\right)  + \frac{1}{2} \ln \left( \frac{\tilde{\sigma}}{\sqrt{\gamma}} \right) \right)d \tp d \tpp\\
\nonumber
&& + \int^{t} \int^{t} (162)^2 H(\tp)^4 H(\tpp)^4 \left( \frac{1}{a(\tp)^2 a(\tpp)^2 \kappa^2 \beta^4}  \right) 
\tilde{\eta} (\tilde{aH})^4 \left[ \left( \frac{1}{2} + \frac{1}{4} ( \ln (\tilde{\sigma}) + 3 \ln \frac{\tilde{\sigma}}{\sqrt{\gamma}} ) \right) 
		+ \left(\frac{k}{\tilde{aH}}\right)^2 \left( \frac{3}{2} \ln \frac{\tilde{\sigma}}{\sqrt{\gamma}} \right) \right . \\
&& \ \ \ \ \ \ \ \ 
\ \ \ \ \ \quad \quad \quad \quad  \left. 
+  \left(\frac{k}{\tilde{aH}}\right)^4 \left( 3  \left( \ln \frac{\tilde{\sigma}}{\sqrt{\gamma}} \right)^2 +  \ln \tilde{\sigma} \ln \frac{\tilde{\sigma}}{\sqrt{\gamma}} 
																		\right) \right] d \tp d \tpp 
\end{eqnarray}
\end{small}
and
\begin{small}
\begin{eqnarray}
\nonumber
&&<\intkp \intkpp {\cal{L}} {\cal{Q}}(\tp, \kp;k) \psikppminusk \psikpp d^3 \vec{\kp} d^3 \vec{\kpp}> = 24 \int^{t} H^2 (H \tp) 
					           \eta \left( 2 (\ln(\sigma))^2 - \frac{3}{2} \ln(\sigma) \ln(\gamma) + \frac{1}{4} \left( \ln(\gamma) \right)^2 \right)d \tp \\
&& \ \ \ \ \ \ \ \ \ \ \ \ - 162  \int^{t} H^2 \frac{1}{a^2 H^2} \frac{H^4}{\kappa \beta^2} 
		\eta (aH)^2 \left( \left[1 + 2 \left(\frac{k}{aH}\right)^2 \ln \left(\frac{\sigma}{\sqrt{\gamma}}\right) \right]
					  \ln  \left( \frac{\sigma^2}{2\sqrt{\gamma}}\right)  + \frac{1}{2} \ln \left( \frac{\sigma}{\sqrt{\gamma}} \right) \right) d \tp,
\end{eqnarray}
\end{small}
where we remind the reader that $\sigma, \gamma, \eta$ are defined in definitions (A13)-(A15) and the $\tilde{}$ symbol denotes symmetrization in $\tp, \tpp$. One can once 
again verify that the dimensions of all of the terms in equations (B9) and (B10) are inverse seconds squared and inverse seconds respectively, as required.

Once again we can immediately take the homogeneous limit directly from the above 
equations (again remembering to add in the contributions from the squares of the averages of the two point functions, as in Appendix A), or by using the relations 
(A16)-(A18) directly in (B7) and (B8). For example, the result for 
$\lim_{k\rightarrow 0} <\intkp \intkpp {\cal{L}} {\cal{Q}}(\tp, \kp;k) \psikppminusk \psikpp d^3 \vec{\kp} d^3 \vec{\kpp}>$ is, putting in the explicit time 
dependence of the background,
\begin{eqnarray}
\lim_{k\rightarrow 0} <\intkp \intkpp {\cal{L}} {\cal{Q}}(\tp, \kp;k) \psikppminusk \psikpp d^3 \vec{\kp} d^3 \vec{\kpp}> &=& 
\left[ 3(24) \frac{ (9 \kappa \pi)^2 }{ (2 \pi)^4} \frac{\alpha^{15}}{\kappa^2 \beta^4} \int^{t} \frac{d \tp}{\tp^{14}} 
											\left( \ln \frac{a_{0} \tp^{\alpha} \alpha}{ \tp k_{min} } \right)^2  \right . \\
\nonumber
&& \left . -3(162) \frac{ (9 \kappa \pi)^2 }{ (2 \pi)^4} \frac{\alpha^{18}}{\kappa \beta^2 \kappa^2 \beta^4} 
					\int^{t}  \frac{d \tp}{\tp^{18}}  \ln \frac{a_{0} \tp^{\alpha} \alpha}{ \tp k_{min} }   \right]
\end{eqnarray}
and similarly for $\lim_{k\rightarrow 0} < \intkp  {\cal{L}} {\cal{Q}}_{\kp - k}  {\cal{L}} {\cal{Q}}_{\kp} d^3 \vec{\kp} >$. 

The key issue is now one of simplifying (B9)-(B10) to extract their dominant parts (those 'most enhanced' by factors of the slow-roll parameter). It should be noted that 
this is complicated by the fact that the dependence on the slow-roll parameter is not just carried through factors of $V^{\prime} \sim \beta$, but also through $\alpha$, which 
in the Introduction we defined to be proportional to $1 / \sqrt{\epsilon_{SR}}$. Therefore we must, to be safe,  carry out the time integrals {\it before} we make the 
approximation that $\alpha >> 1, \epsilon_{SR} << 1, \alpha / t \rightarrow H \in \Re$ epsecially since, just as differentiation in comoving time of the scale factor generally 
introduces factors of the slow-roll parameter in the numerator of a given expression, integration introduces factors in the denominator. This is the reason it is not 
immediately obvious the second term in equation (B2) is subdominant to the first term. 

We begin in the simpler case of the homogeneous limit. In that case, one can for example show that the dominant terms for the following integrals are (for $\alpha >> 1$):
\begin{eqnarray}
\int^{t} \frac{d \tp}{\tp^{14}} \left( \ln \frac{a_{0} \tp^{\alpha} \alpha}{ \tp k_{min} } \right)^2 &\approx&
			-\frac{2 \alpha^2}{2197 t^{13}} - \frac{2\alpha}{2197 t^{13}} \left( 13 \ln{\sigma} -2 \right)\\
\int^{t} \frac{d \tp}{\tp^{18}} \left( \ln \frac{a_{0} \tp^{\alpha} \alpha}{ \tp k_{min} } \right) &\approx&
			-\frac{ \alpha}{289 t^{17}} + \frac{1}{289 t^{17}} \left( -17\ln{\sigma} + 1 \right) 
\end{eqnarray}
and similarly for the other integrals required. Using these sorts of integrals in equations (B11) for example, one can show that
\begin{small}
\begin{eqnarray}
\lim_{k\rightarrow 0} <\intkp \intkpp {\cal{L}} {\cal{Q}}(\tp, \kp;k) \psikppminusk \psikpp d^3 \vec{\kp} d^3 \vec{\kpp}> &=&
-3(162) \frac{ (9 \kappa \pi)^2 }{ (2 \pi)^4} \frac{H^5 }{ 289 {\epsilon_{SR}}^3 } \left[ -\alpha^2 + \alpha (-17 \ln( \sigma) ) \right], 
\end{eqnarray}
\end{small}
and similarly for $\lim_{k\rightarrow 0} < \intkp  {\cal{L}} {\cal{Q}}_{\kp - k}  {\cal{L}} {\cal{Q}}_{\kp} d^3 \vec{\kp} >$.

For the full inhomogeneous problem we can make excellent use of the long-wavelength approximation to rewrite equations (B9) and (B10) as
\begin{small}
\begin{eqnarray}
\nonumber
< \intkp  {\cal{L}} {\cal{Q}}_{\kp - k}&&{\cal{L}} {\cal{Q}}_{\kp} d^3 \vec{\kp} > = 
\int^{t} \int^{t} H(\tp)^3 H(\tpp)^3 (24)^2 \tp \tpp  
				\tilde{\eta} \left( 2 (\ln(\tilde{\sigma}))^2 - \frac{3}{2} \ln(\tilde{\sigma}) \ln(\gamma) 
										+ \frac{1}{4} \left( \ln(\gamma) \right)^2 \right) d \tp d\tpp \\
&& -  \int^{t} \int^{t} H(\tp)^4 H(\tpp)^3 (24 \tp) \left( \frac{162}{a(\tp)^2 \kappa \beta^2}  \right)  \tilde{\eta} (\tilde{aH})^2 \left( 
		\ln  \left( \frac{{\tilde{\sigma}}^2}{2\sqrt{\gamma}}\right)  + \frac{1}{2} \ln \left( \frac{\tilde{\sigma}}{\sqrt{\gamma}} \right) \right)d \tp d \tpp\\
\nonumber
&& -  \int^{t} \int^{t} H(\tpp)^4 H(\tp)^3 (24 \tpp) \left( \frac{162}{a(\tpp)^2 \kappa \beta^2}  \right) \tilde{\eta} (\tilde{aH})^2 \left( 
	\ln  \left( \frac{{\tilde{\sigma}}^2}{2\sqrt{\gamma}}\right)  + \frac{1}{2} \ln \left( \frac{\tilde{\sigma}}{\sqrt{\gamma}} \right) \right)d \tp d \tpp\\
\nonumber
&& + \int^{t} \int^{t} (162)^2 H(\tp)^4 H(\tpp)^4 \left( \frac{1}{a(\tp)^2 a(\tpp)^2 \kappa^2 \beta^4}  \right) 
\tilde{\eta} (\tilde{aH})^4 \left[ \left( \frac{1}{2} + \frac{1}{4} ( \ln (\tilde{\sigma}) + 3 \ln \frac{\tilde{\sigma}}{\sqrt{\gamma}} ) \right) \right] 
																			 d \tp d \tpp 
\end{eqnarray}
\end{small}
and
\begin{small}
\begin{eqnarray}
\nonumber
<\intkp \intkpp {\cal{L}} {\cal{Q}}(\tp, \kp;k) \psikppminusk \psikpp d^3 \vec{\kp} d^3 \vec{\kpp}> &\approx& 24 \int^{t} H^2 (H \tp) 
					           \eta \left( 2 (\ln(\sigma))^2 - \frac{3}{2} \ln(\sigma) \ln(\gamma) + \frac{1}{4} \left( \ln(\gamma) \right)^2 \right)d \tp \\
&& - 162  \int^{t} H^2 \frac{H^4}{\kappa \beta^2} \eta \left( 
					\ln  \left( \frac{\sigma^2}{2\sqrt{\gamma}}\right)  + \frac{1}{2} \ln \left( \frac{\sigma}{\sqrt{\gamma}} \right) \right) d \tp
\end{eqnarray}
\end{small}
It turns out the dominant terms in these integrals are, for $\alpha >> 1$ and $\epsilon_{SR} << 1$,  
\begin{small}
\begin{eqnarray}
< \intkp  {\cal{L}} {\cal{Q}}_{\kp - k}  {\cal{L}} {\cal{Q}}_{\kp} d^3 \vec{\kp} > &\approx& 
-\frac{531441 \kappa^2 H^6}{360448 \pi^2 { \epsilon_{SR} }^4 } \left[ 192 \alpha + 7040 \ln (\sigma) - 2112 \ln (\gamma) \right]  \\
<\intkp \intkpp {\cal{L}} {\cal{Q}}(\tp, \kp;k) \psikppminusk \psikpp d^3 \vec{\kp} d^3 \vec{\kpp}> &\approx&
\frac{\kappa^2 H^5}{4624 \pi^2 {\epsilon_{SR} }^{3}} \left[ 32805 \alpha^2 -  6561 \alpha (-85 \ln (\sigma) + 51 \ln( \gamma) ) \right], 
\end{eqnarray}
\end{small}
where we note that $k$ dependence comes in solely from the $\ln (\gamma)$ terms and that the dimensions of equations (B17) and (B18) are respectively inverse seconds squared 
and inverse seconds, as required. 

We are now finally in a position to collect all of these results, namely, equations 
(B17)-(B18) and (A10) (and (B15)-(B16), (A16)), to fully evaluate  the quantity $\sqrt{< \left( \frac{ \delta^2 \rho_{IR} }{\rhobar} \right)^2 >}$ at some scale 
$\kt$ such that $k_{min} < \kt << aH$ (and also at $\kt=0$). The goal is to compare the magnitude of this term to the horizon-scale amplitude of the linearized fluctuations   
$\sqrt{< \left . \left( \frac{ \delta \rho }{\rhobar} \right)^2 \right|_{k=aH} >}$, as shown in Appendix C.

\section{Derivation of Inequalities (60) and (73) }

In order to derive inequalities (60) and (73) we must calculate the VEV amplitude of the second order fluctuations at the homogeneous scale. We shall derive the result for 
some scale $k_{min} \sim \kt << aH$ and then take the homogeneous limit to recover inequality (60). 
Using the results and notation of Appendices A and B on the dominant (scalar-scalar) terms of equation (50), one may write the averaged square of the dominant contributions 
to the second order IR pressure contribution (or energy density, etc.) as  
\begin{eqnarray}
\nonumber
< \delta^2 p_{IR}(k) \delta^2 p^{\dagger}_{IR}(k) > &\approx& 
			< \intkp \intkpp \left( \frac{ 54 H^2}{\kappa \epsilon_{SR}} \right)^2 \psikpminusk \psikp \psikppminusk \psikpp d^3 \vec{\kp} d^3 \vec{\kpp} > \\
\nonumber
&+& <\intkp \left( \frac{ 3H}{\kappa} \right)^2 {\cal{L}} {\cal{Q}}_{\kp - k}  {\cal{L}} {\cal{Q}}_{\kp} d^3 \vec{\kp}  > \\
&+& 
< \intkp \intkpp \left( \frac{ 6H}{\kappa}  \frac{ 54 H^2}{\kappa \epsilon_{SR}} \right) {\cal{L}} {\cal{Q}}(\tp, \kp;k) \psikppminusk \psikpp d^3 \vec{\kp} d^3 \vec{\kpp}   >
\end{eqnarray} 
Putting in the results from all of the above appendices we find that the dominant terms are of the form 
\begin{eqnarray}
< \delta^2 p_{IR}(k) \delta^2 p^{\dagger}_{IR}(k) > &\approx& 
\left( \frac{H^2}{\kappa} \right)^2 \frac{\kappa^2 H^4}{\epsilon_{SR}^4 \pi^2} \left( A_{1} \alpha^2 + \alpha \left( B_{1} \ln (\gamma) + C_{1} \ln (\sigma) \right) \right), 
\end{eqnarray}
where 
\begin{eqnarray}
A_{1} &\equiv& \frac{2657205}{1156 \pi^2} \\
B_{1} &\equiv& \frac{-1594323}{68 \pi^2 } \\
C_{1} &\equiv& \frac{2657205}{68 \pi^2}
\end{eqnarray}
It is important to note that the details of this calculation confirm that the naive guess ventured in the Introduction holds: there are solutions of the diffeomorphism 
constraints of general relativity,  at second order in perturbation theory about a slowly rolling background, 
which introduce a factor of $1/\epsilon_{SR}$ into any expression of $\delta^2 \rho_{IR}$ and $\delta^2 p_{IR}$, and these factors survive through quantum averaging. 
A second order gauge transformation cannot eliminate the presence of such slow-roll enhanced terms.  

The central result of this paper is in comparing the amplitudes of the second order fluctuations at $k_{min} < k = \kt << aH$ to the amplitudes 
linearized fluctuations at the horizon scale, $k = aH$. Using equation (C2) and equation (55) for this purpose, we see that 
\begin{eqnarray}
\nonumber
\sqrt{< \left . \left( \frac{ \delta p }{\pbar} \right)^2 \right|_{k=aH} >} \  &>& \   \left . \sqrt{< \left( \frac{ \delta^2 p_{IR} }{\pbar} \right)^2 >} \right|_{k = \kt}
\end{eqnarray}
is equivalent to demanding that
\begin{eqnarray}
\frac{\tilde{A}_{2} \sqrt{\kappa} H}{\sqrt{\epsilon_{SR} \pi } }&>& 
\frac{\kappa H^2}{ {\epsilon_{SR}}^{2} \pi^2 } \sqrt{A_{1} \frac{N}{\epsilon_{SR}}  + \sqrt{\frac{N}{{\epsilon_{SR}}}} ( B_{1} \ln (\gamma) + C_{1} \ln (\sigma) ) }, 
\end{eqnarray}
where $\alpha^2 = N / \epsilon_{SR}$, $\tilde{A}_{2} = \frac{9}{16}$. We can see that the spatial dependence is subdominant in the sense that it is multiplied by a lower 
power of $\alpha$: in other words, the homogeneous fluctuations (mean) are not, in the slow-roll limit, altered by the finite corrections which occur when 
one evaluated at the finite spatial scale of $k = \kt << aH$. We also note that (C6) is positive definite, as it should be, since $k << aH$. Furthermore we note that 
the first and second order dispersions above have the same weighting in volume normalization factors, so that they cancel. Thus, we find that 
\begin{eqnarray}
\epsilon_{SR} > \frac{2}{3} ( \kappa H^2 )^{\frac{1}{4}} {(A_{1}N) }^{\frac{1}{4}}
\end{eqnarray}
is the consistency condition for linearized theory at second order.

\section{ Scalar metric fluctuations in de-Sitter }

In this section we comment that to second order in perturbation theory about pure de-Sitter (no-roll), the scalar sector is nontrivial. This can also be seen by e.g. 
examining the second order gauge fixing (45) in the main paper or the reduced equation of motion for ${\cal{Q}}$, equation (48), which shows that the TT sector will mix with 
and source a nontrivial scalar sector at second order when $\epsilon_{SR} \rightarrow 0$. It is perhaps worth emphasizing that the extra metric functions 
one can fix in flat, vacuum, spacetime and special spacetimes like de Sitter are generic, but obey field equations. Though they correspond to residual degrees of freedom
specified only on an initial value surface, they are completely determined by equations of motion they {\it also} satisfy. Therefore any residual gauge fixing makes 
crucial use of the equations of motion to set additional terms to zero via a gauge transformation.

To provide a formal alternate argument which however is far more compact and hopefully more convincing, consider the following.
To linear order in perturbation theory about pure vacuum de-Sitter it is relatively straightforward that one can gauge away the scalar sector. Let us go through the 
proof and then broadly identify the impediment which would provent extending this procedure to second order. 

One may compactly prove this by writing by linearizing the linearized Einstein equations about 
\begin{eqnarray}
d\bar{s}^{2} &=& \bar{g}_{a b} d \bar{x}^{a} d \bar{x}^{b} = -dt^{2} + \cosh(t)^{2} 
                                 \left[ d \chi^{2} + \sin^{2}(\chi) d \Omega^{2} \right],
\end{eqnarray}
which yields 
\begin{eqnarray}
\bbbox \delta g_{ab} - \frac{1}{2} \bar{g}_{ab} \bbbox \delta g - 2 \delta g_{ab} - \bar{g}_{ab} \delta g &=& 0
\end{eqnarray}
provided we choose the transverse Lorentz gauge, i.e. 
\begin{eqnarray}
\bar{\nabla}^{a} \left[ \delta g_{a b} - \frac{ \bar{g}_{a b}}{2} \delta g \right] &=& 0.
\end{eqnarray}
One then asks if one can impose tracelessness as well, and it turns out that making the additional choice 
$\delta g^{\prime}_{a b} \equiv \delta g_{a b} + \frac{1}{6} \bar{\nabla}_{a} \bar{\nabla}_{b} \delta g$ is both consistent with the imposition of the Lorentz gauge (D3)
and also satisfies the field equations (D2).  However, the imposition of the Lorentz gauge along with tracelessness still leaves a residual gauge freedom in the form of 
gauge transformations which are harmonic functions of the operator $\stackrel{\bar{{}^{}}}{\Box} + \Lambda$, with which we can hope to eliminate $\delta g^{\prime}_{0i}$. Not 
only does it turn that such transformations are possible, but  
the field equations expressed in this new gauge (combined with the maximal symmetry of de Sitter) also demand that $\delta g^{\prime}_{00} = 0$. In the complete, closed,  
slicing (D1) the fact that the lapse fluctuation $\delta g^{\prime}_{00}$ is zero corresponds to the fact that the eigenfunctions of the Laplacian on $S^{3}$ must be 
periodic (or, in other words, their eigenspectrum must be discrete). In other words, a positivity argument shows that the linear lapse fluctuation can {\it also} 
be gauged away. Thus tracelessness, transverseness, and the vanishing of the shift and lapse leaves two degrees of freedom, which are the two polarizations of the graviton.

Therefore, at linear order in perturbation theory, the scalar sector is trivial (as is the vector--the proof is considerably simpler) modulo a linear gauge transformation 
and only the TT sector remains, which to this order is gauge invariant anyway. 

At second order, however, additional gravitational wave terms in TT-TT combinations (which transform as scalars)
will act as sources for the scalar perturbations. These additional terms will enter as both negative definite and positive definite contributions because of the form of the 
perturbed Christoffel symbols (whose quadratic products and derivatives comprise the field equations), and this variable sign contribution is sufficient  
to in general get a nontrivial solution for the second order lapse. Therefore it is not possible in general to gauge it away at second order if one also insists on gauging 
away the second order shifts, so the second order scalar sector is not a gauge artifact.

\end{document}